\documentclass{article}
\usepackage{amsfonts,amsmath,epsf}
\usepackage{graphics} 

\advance\voffset by -1.6cm
\advance\hoffset by -1.7cm
\newcounter{defthm}
 
\newtheorem{defthm}{\whattheorem}[section]

\newcommand\sect[1]{\section{#1}\setcounter{equation}0\setcounter{defthm}0}

\newcommand\void[1]       {}

\newcommand\be            {\begin{equation}}
\newcommand\bea           {\begin{equation}\begin{array}l\displaystyle}
\newcommand\bearll        {\begin{array}{ll}\displaystyle}
\newcommand\ee            {\end{equation}}
\newcommand\eear          {\end{array}}
\newcommand\enl           {\\[1em]\displaystyle}
\newcommand\etb           {& \displaystyle}
\newcommand\erf[1] {(\ref{#1})}
\newcommand\labl[1]       {\label{#1}\ee}

\newcommand\eps           {\varepsilon}

\newcommand\mem           {\hspace*{-0.5em}}
\newcommand\one           {{\bf1}}

\newcommand\Cb            {\mathbb{C}}
\newcommand\Rb            {\mathbb{R}}
\newcommand\Zb            {\mathbb{Z}}

\newcommand\Cc            {\mathcal{C}}

\def\thefootnote{\fnsymbol{footnote}}
\begin{document}

\begin{flushright}  {~} \\[-12mm]
{\sf hep-th/0509040}\\[1mm]
{\sf AEI-2005-139}\\[1mm]
{\sf MAD-TH-05-6}\\[1mm]
\end{flushright} 

\thispagestyle{empty}

\begin{center} \vskip 14mm
{\Large\bf Matrix model eigenvalue integrals and\\[0.5em]
   twist fields in the $su(2)$-WZW model}\\[20mm] 
{\large 
Matthias R.\ Gaberdiel\,${}^1$,\,  
Albrecht Klemm\,${}^2$,\, 
Ingo Runkel\,${}^3$\footnote{{\tt Emails: gaberdiel@itp.phys.ethz.ch,
aklemm@physics.wisc.edu, ingo@aei.mpg.de}}}
\\[8mm]
${}^1$ Institut f\"ur Theoretische Physik, ETH Z\"urich, Switzerland\\[1mm]
${}^2$ UW-Madison Physics Dept, Madison, USA\\[1mm]
${}^3$ Max-Planck-Institut f\"ur Gravitationsphysik, Potsdam, Germany
\end{center}
\vskip 22mm

\begin{quote}{\bf Abstract}\\[1mm]
We propose a formula for the eigenvalue integral of the
hermitian one matrix model with infinite well potential in terms of
dressed twist fields of the $su(2)$ level one WZW model. The
expression holds for arbitrary matrix size $n$, and provides a
suggestive interpretation for the monodromy properties of the matrix
model correlators at finite $n$, as well as in the $1/n$-expansion. 
\end{quote}
\vfill
\newpage 

\setcounter{footnote}{0}
\def\thefootnote{\arabic{footnote}}

\sect{Introduction}

It has been known for some time that there is a close relation between
the hermitian one matrix model and the conformal field theory of one
free boson 
\cite{Marshakov:1991gc,Kharchev:1992iv, Morozov:hh, Kostov:1999xi}. 
One of the quantities that appears naturally in the matrix model context
corresponds, in conformal field theory, to the operator 
\be
  I_\lambda(a,b) = \exp\!\Big( \frac{\lambda}{2\pi}
  \int_a^b \!\!J^+(x) \, dx \Big) \,.
\labl{eq:Idef}
Here $\lambda$ is a complex number and we have expressed the free
boson theory (at the self-dual radius) in terms of the $su(2)$ level
one WZW model with the standard basis of the weight one fields denoted
by $J^+(z)$, $J^3(z)$ and $J^-(z)$. Since the operator product
expansion of $J^+(z) J^+(w)$ is regular, no normal ordering is
required to ensure convergence of the integrals 
that occur upon expanding \erf{eq:Idef}.

As will be explained in more detail in the next subsection, the matrix 
model analysis suggests that the operator $I_\lambda(a,b)$ will
exhibit interesting monodromy properties for the $su(2)$ fields in the
limit of large matrix size $n$. The aim of this paper is to elucidate
these in terms of the above conformal field theory description. In 
particular, we shall propose that the operator \erf{eq:Idef} can also
be expressed in terms of dressed twist fields (see section
\ref{sec:summary} below for more details) which will make these
monodromy properties manifest. To leading order in $1/n$ the relation
between $I_\lambda(a,b)$ and twist fields had been suggested before in  
\cite{Kostov:1999xi,Dijkgraaf:2002yn}; here we shall propose an exact 
relation for all $n$.

\subsection{Relation to matrix models and 2d gravity}\label{sec:mat-2dgrav}

The matrix integral for the hermitian one matrix model
is given by
\be
Z_{\rm mm}\big[{\underline t}\big]^{(n)}
   =~ ({\rm const}) \int\!\!d\Phi\,e^{-\frac{1}{g_s} {\rm tr} W(\Phi) }
   ~= \int_{-\infty}^\infty \!\!\!\!\! d\lambda_1 \dots d\lambda_n
    \!\!\! \prod_{i,j=1,\,i<j}^n \!\!\! (\lambda_i - \lambda_j)^2 ~
    e^{-\frac{1}{g_s} \sum_{k=1}^n W(\lambda_k)} \,.
\labl{eq:herm-mat}
The first integral is over all hermitian $n{\times}n$ matrices,
while the second integral amounts to expressing the first one in
terms of the eigenvalues of $\Phi$, 
and $W(x)=\sum_{m\ge 0} t_m x^m$ is the potential;  
more details can be found {\it e.g.}\ in the review \cite{DiFrancesco:1993nw}. 
The relation to the free boson conformal field theory is established
by noting that the integrand of \erf{eq:herm-mat} can be written
as a correlator of free boson vertex operators
\cite{Marshakov:1991gc,Kharchev:1992iv, Morozov:hh, Kostov:1999xi}. 
Using the language of $su(2)_1$ the expression is,
\be
  \langle n | e^{-H} J^+(x_1) \cdots J^+(x_n) |0\rangle
  = e^{-\frac{1}{g_s} \sum_{k=1}^n W(x_k)}
    \prod_{i<j} (x_i - x_j)^2  
  \,,
\labl{eq:integrand-as-corr}
where $H = \tfrac{1}{g_s} 
\int_{\Cc_\infty} W(z) J^3(z) \tfrac{dz}{2\pi i }$.
This formula is most easily verified by employing the operator product 
expansion $J^3(z)J^+(w) = (z{-}w)^{-1} J^+(w) + O(1)$ to commute 
$e^{-H}$ to the right. 
Here we have assumed that $W(x)$ 
is analytic on $\Cb$: this allows us
to use contour integrals, and guarantees that the only contributions
come from the $J^+$-insertions. 
Using charge conservation, the matrix integral can then be expressed 
in terms of the exponential \erf{eq:Idef} as
\be
  \langle n | e^{-H} I_\lambda(-\infty,\infty) |0\rangle
  =  \frac{\lambda^{n}}{(2\pi)^n \, n!} Z_{\rm mm}\big[\underline t\big]^{(n)} \,.
\labl{eq:Zmm=Zcft}

The correspondence extends also to correlators. 
Let $Z_{\rm mm}[{\underline t},f]^{(n)}$ stand for the the right hand side of 
\erf{eq:herm-mat}, with an additional factor $f$ in the integrand. 
For the matrix model resolvent we then have the relation
\be
  \langle n | e^{-H} J^3(z) I_\lambda(-\infty,\infty) |0\rangle
  = \frac{\lambda^{n}}{(2\pi)^n \, n!}
  Z_{\rm mm}\big[\, {\underline t},\,  {\rm tr}\tfrac{1}{z-M} - 
  \tfrac{1}{2g_s} W'(z) \big]^{(n)} \,,
\labl{eq:resolv-rel}
as can be verified by writing out the integrals explicitly and
comparing the integrands.

Provided the potential $W(x)$ increases fast enough for 
$x \rightarrow \pm\infty$, the eigenvalues of the matrix $\Phi$ will
condense, in the large $n$ limit, on one or more intervals on the 
real axis, the so-called cuts 
(for details, consult {\it e.g.}\ \cite{DiFrancesco:1993nw}).
The correlator $Z_{\rm mm}\big[{\underline t}, {\rm tr}\tfrac{1}{z-M} \big]^{(n)}$ has,
again in the large $n$ limit, square root branch cuts on these
intervals. Equivalently, continuing $J^3(z)$ in \erf{eq:resolv-rel}
around an endpoint of a cut in the large $n$ limit
results in $J^3 \rightarrow -J^3$. 
This motivates the idea to model the endpoints of matrix model cuts
by insertions of twist fields $\sigma(x)$ on the conformal field
theory side \cite{Kostov:1999xi,Dijkgraaf:2002yn}.
Free boson twist fields have also been considered in the context
of the integrable hierarchy approach to two-dimensional gravity in
\cite{Dijkgraaf:1990rs, 
Fukuma:1990jw, Fukuma:1996hj}.

We will investigate this effect in a simplified setting, in which we
choose $W(x)$ in \erf{eq:herm-mat} to be an infinite well potential, 
that is, $W(x)=0$ for $x \in [a,b]$ and $W(x) = +\infty$ otherwise. 
This effectively restricts the eigenvalue integrations from the
real axis to the interval $[a,b]$, so that the relation
\erf{eq:Zmm=Zcft} takes the simpler form
\be
  \langle n | I_\lambda(a,b) |0\rangle 
  = \frac{ \lambda^n }{ (2\pi)^n \, n! } ~ Z_{\rm mm}^{\rm well,(n)} 
  \,.
\labl{eq:Zmm-Zcft-simple}
Matrix integration measures (or potentials) which force the
eigenvalues to lie on a contour which has an endpoint on
the complex plane, rather then to start and end at infinity,
are referred to as ensembles with hard edges, see 
{\it e.g.}\ \cite{Borodin:2003,Eynard:2005rg} 
where such models are treated and more references
can be found.
{}From the conformal field theory point of view, the relation
\erf{eq:Zmm-Zcft-simple} is easier to analyse than \erf{eq:Zmm=Zcft}
because the potential term $e^{-H}$ is absent. For example, for the
one-point functions one finds
\be\bearll
  \langle n | J^3(z) I_\lambda(a,b) |0\rangle
  \etb = \frac{ \lambda^n }{ (2\pi)^n \, n! } ~  
  Z_{\rm mm}^{\rm well}\big[ {\rm tr}\tfrac{1}{z-M}  \big]^{(n)} 
  \enl
  \langle n{\pm}1 | J^\pm(z) I_\lambda(a,b) |0\rangle
  \etb = \frac{ \lambda^n }{ (2\pi)^n \, n! } ~ 
  Z_{\rm mm}^{\rm well}\big[ {\rm det}(z-M)^{\pm 2}  \big]^{(n)}\,.
\eear\labl{eq:mm-well-1pt}
It should be stressed that even if we
choose the interval $[a,b]$ to coincide with the location of a cut in 
a matrix model with analytic potential $W(x)$,
the large $n$ expansion of the free energies 
\be
  \ln\!\big( \langle n | e^{-H} J^3(z) I_\lambda(a,b) |0\rangle \big)
  \quad {\rm and} \quad
  \ln\!\big( \langle n | e^{-H} J^3(z) I_\lambda(-\infty,\infty) 
  |0\rangle \big) ~,
\ee
as well as
their behaviour in the double scaling limit, will be different.
The reason is that for subleading effects in $n^{-1}$, the 
decay-behaviour of the eigenvalues just outside of the cut
will be important, and that has precisely been cut off by
the infinite well potential\footnote{We thank B.~Eynard for a
discussion on this point.}.  

\medskip

As we have explained above one may expect, given the monodromy
properties of the matrix model correlators, that the operator  
$I_\lambda(a,b)$ is proportional to the product of two twist fields 
to leading order in $1/n$. For finite $n$, the monodromy properties
are  however quite different. Indeed, the correlator   
$Z_{\rm mm}[{\underline t},{\rm det}(z-M) ]^{(n)}$ is single 
valued for finite $n$
(being a polynomial of degree $n$), but has jumps at the location of
the cuts in the $1/n$-expansion. This is an example of Stokes'
phenomenon (for an exposition see for example
\cite{Berry}, \cite[app.\,B]{Maldacena:2004sn}): the analytic continuation 
in $z$ of an asymptotic expansion (the $1/n$-expansion in this case)
of a function $f(n,z)$ can be different from the asymptotic expansion
of the analytic continuation. 

The double scaling limit of the hermitian matrix model, with an
appropriately tuned potential, describes a $(p,2)$-minimal
model coupled to Liouville gravity (see {\it e.g.}\ 
\cite{DiFrancesco:1993nw}) 
or, equivalently, $(p,2)$-minimal string theory
(see \cite{Seiberg:2004at} for a summary of recent developments).
The above effect has been given an interesting target space 
interpretation
in the context of minimal string theory \cite{Maldacena:2004sn}.
There, the target space is identified with the moduli space
of FZZT branes and it is shown that while 
in the semi-classical limit this moduli space becomes a 
branched covering of the complex plane, in the exact quantum
description the moduli space is in fact much simpler, being
just the complex plane itself. 
On the matrix model side, the analogue of the
string partition function in the
presence of a FZZT brane is the correlator of the exponentiated
macroscopic loop operator, 
$Z_{\rm mm}[{\underline t},  {\rm det}(z-M) ]^{(n)}$  
\cite{Banks:1989df,Moore:1991ir,Moore:1991ag,Kostov:1991hn}. 
As described above, this correlator is single valued for finite $n$
but develops branch cuts in the large $n$ limit.
This is an example for how classical geometry emerges as an effective
concept in string theory.

As we will see, also the simplified matrix model with the
infinite well potential that we consider in this paper exhibits this
behaviour. The alternative formula for $I_\lambda(a,b)$ in terms of
dressed twist fields that we shall propose will then give a suggestive
explanation of this phenomenon.

\subsection{Summary of results}\label{sec:summary}

To explain more precisely our formula for $I_\lambda(a,b)$ in terms of 
dressed twist fields, we first need to give a few definitions, 
starting with the
relevant twist fields $\sigma_{\pm\lambda}(z)$. Around an insertion of  
$\sigma_{\pm\lambda}(z)$, the three $su(2)$-currents have a
$\Zb_2$-monodromy
\be
  (J^+,J^3,J^-) \longmapsto 
  (\lambda^{-2} J^-, -J^3, \lambda^{2} J^+)\,.
\labl{eq:Z2-mono}
If instead of $J^c$ one uses the basis
\be
  K^3 = \tfrac12\big(\lambda J^+ + \lambda^{-1} J^-\big) \,,
\qquad
  K^\nu = \tfrac{\nu}{2}\big(\lambda J^+ - \lambda^{-1} J^-\big) - J^3
  \quad {\rm for} \, \nu = \pm 1 ~,
\labl{eq:K-basis}
the twist fields $\sigma_{\pm\lambda}(z)$ have the standard monodromy
$K^3 \mapsto K^3$ and $K^\nu \mapsto - K^\nu$. In particular, the
field $K^3(z)$ is single valued, and the two twist fields 
$\sigma_{+\lambda}(z)$ and $\sigma_{-\lambda}(z)$ are distinguished
by their $K^3$-charge ({\it i.e.}\ the eigenvalue of the $K^3$ zero mode), 
which is $\tfrac14$ and $-\tfrac14$, respectively.
In the $K$-basis it is obvious that the monodromy around
$\sigma_{\pm\lambda}(z)$ amounts to an inner automorphism of
$su(2)$ of order two.

Let us define an operator $S_\lambda(a,b)$ which is very similar
in spirit to (a special case of) the star-operators introduced
in \cite{Moore:1990mg,Moore:1990cn}. Explicitly, it is given
in terms of twist fields and exponentiated $J^-$-integrals as
follows,
\be
  S_\lambda(a,b) = (b-a)^{\frac18}
  \Big[ \sigma_{+\lambda}(b)
  \exp\Big( - \frac{1}{2\pi\lambda}
  \int_{\Cc_1} J^-(x) dx \Big) 
  \exp\Big( - \frac{1}{2\pi\lambda}
  \int_{\Cc_2} J^-(x) dx \Big) 
    \sigma_{-\lambda}(a) \Big]_{\rm reg} \,.
\labl{eq:Sdef}
Here $\Cc_1$ is an integration contour from $a$ to $b$
passing above the interval $[a,b]$, the contour $\Cc_2$
has the same endpoints, but passes below the interval,
and $[\cdots]_{\rm reg}$ refers to a prescription to 
regulate the first order pole in the operator product expansion of
$J^-$ and $\sigma_{\pm\lambda}$ (see sections \ref{sec:regulate}  
and \ref{sec:S-norm} below). 
The contours are illustrated in the following figure,
\be
  \begin{picture}(300,70)(0,0)
  \put(0,0){\scalebox{0.5}{\includegraphics{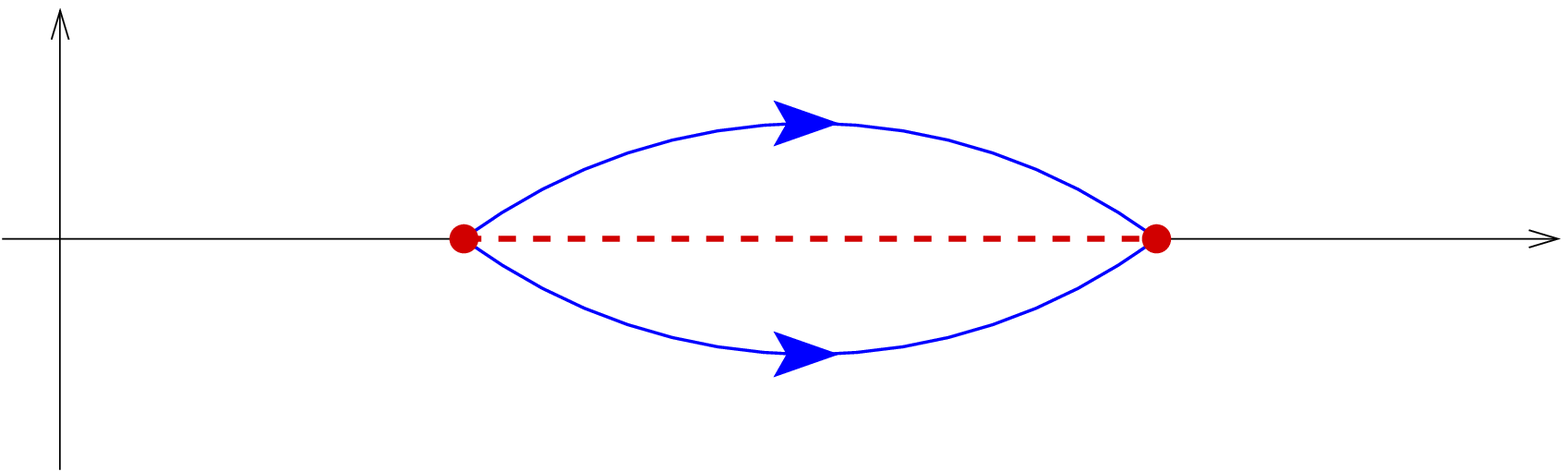}}}
  \put(42,27){$\sigma_{-\lambda}(a)$}
  \put(180,27){$\sigma_{\lambda}(b)$}
  \put(110,58){$\Cc_1$}
  \put(110,10){$\Cc_2$}
  \put(13,60){Im}
  \put(225,27){Re}
  \end{picture}
\labl{eq:contour-lens}
The dashed line represents the branch cut between the two
twist fields.

As we shall explain below in section \ref{sec:twistrep}, the 
product of the two twist fields 
$\sigma_{+\lambda}(b) \sigma_{-\lambda}(a)$ that appears in 
$S_\lambda(a,b)$ can be expressed in terms 
of an exponentiated integral of the form 
\be
\sigma_{+\lambda}(b) \, \sigma_{-\lambda}(a) \,=\,
               (b-a)^{-\frac18}\,
: \exp\!\Bigg(\frac{1}{4} 
\int_a^b \Big( \lambda \, J^+(z) + \lambda^{-1} J^-(z)\Big) \Bigg) : \,.
\ee
Qualitatively speaking, the $J^-$ integrals in the formula
\erf{eq:Sdef} for $S_\lambda(a,b)$ can be interpreted as removing the
$J^-$ part of this integral, leaving behind only the $J^+$ integrals
that appear in $I_\lambda(a,b)$. This observation motivates the
following operator identity for the two rather different looking
exponentiated integrals \erf{eq:Idef} and \erf{eq:Sdef},
\be
   I_\lambda(a,b) = S_{\lambda}(a,b) \,.
\labl{eq:main}
This equality is the main result of our paper. 
We have no complete proof for it; the supporting
evidence will be given in section \ref{sec:Slam}.
\smallskip

As a consequence of \erf{eq:main} it is now possible to see that
the operator $I_\lambda(a,b)$ does indeed display the monodromy
properties described in the previous section.
Consider a correlator of the form
$\langle n | ({\rm fields}) S_\lambda(a,b) |0\rangle$,
where $({\rm fields})$ stands for any product of the currents
$J^c(z)$. The integration contour \erf{eq:contour-lens} can
be deformed as in figure i) below.
\be
  \begin{picture}(350,45)(0,0)
  \put(0,0){
    \put(0,0){\scalebox{0.35}{\includegraphics{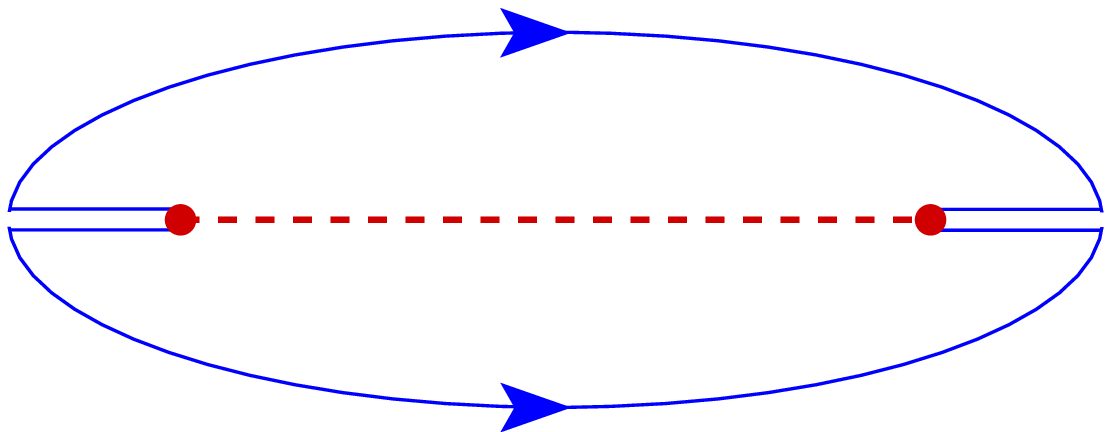}}}
    \put(-10,40){i)}
    \put(15,14){$a$}
    \put(92,12){$b$}
  }
  \put(140,21){
    \put(0,0){\vector(1,0){60}}
    \put(10,3){truncate}
    }
  \put(230,0){
    \put(0,20){\scalebox{0.35}{\includegraphics{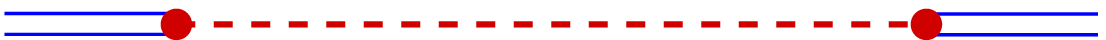}}}
    \put(-10,40){ii)}
    \put(15,14){$a$}
    \put(92,12){$b$}
  }
  \end{picture}
\labl{eq:contour-truncate}
It turns out (see section \ref{sec:non-pert})
that the part of the integral along the
ellipse is suppressed by a factor $r^{-2n}$ for
some $r>0$. This results in the approximation
\be
  \langle n | ({\rm fields}) S_\lambda(a,b) |0\rangle
  =
  \langle n | ({\rm fields}) S^{\rm trunc}_\lambda(a,b) |0\rangle
  \big( 1 + O(r^{-2n}) \big) \,,
\label{eq:S-trunc}
\ee
where $S^{\rm trunc}_\lambda(a,b)$ is defined as $S_\lambda(a,b)$,
but with the $J^-$-integrals taken only over the short horizontal
contours shown in figure 
(\ref{eq:contour-truncate}\,ii).\footnote{As explained in section
\ref{sec:non-pert}, in addition the regulator introduces an overall
factor.} The difference between $S_\lambda(a,b)$ and 
$S^{\rm trunc}_\lambda(a,b)$  
in a correlator with an out-state of charge $n$ is thus non-perturbative
in $1/n$. In particular, both correlators in \erf{eq:S-trunc} will
have the same $1/n$ expansion 
(the correlators have to be normalised appropriately to allow a
$1/n$-expansion, see section \ref{sec:non-pert}).
But since in $S^{\rm trunc}_\lambda(a,b)$ the points $a$ and $b$ are
no longer connected by $J^-$ integrals, the monodromy of the currents
$J^c(z)$ around the points $a$ and $b$ is just the $\Zb_2$-monodromy
\erf{eq:Z2-mono} of the twist fields. On the other hand, the 
monodromy of $I_\lambda(a,b)$ (and hence that of $S_\lambda(a,b)$)
is not given by \erf{eq:Z2-mono}. For example, in the presence of
$I_\lambda(a,b)$ the current $J^+(z)$ is single valued (since the
operator product expansion $J^+(z)J^+(w)$ is regular), while under
\erf{eq:Z2-mono} it changes to $\lambda^{-2} J^-(z)$. 
In this sense,
the `correct' monodromy of the currents $J^c(z)$ in the
presence of $I_\lambda(a,b)$ is a non-perturbative effect and
cannot be seen in the $1/n$-expansion.
This is the same effect as
observed in the previous section for the minimal string, albeit 
here in a different model, namely a matrix integral with hard
edges.

Further, in the large $n$ limit itself, the $J^-$-integrals are
suppressed altogether, and $S_\lambda(a,b)$ can be replaced by
a product of twist fields. 
Taking into account the need to regulate \erf{eq:Sdef} 
(see section \ref{sec:non-pert}), we arrive in this way at the
second important result of our paper,
\be
  \langle n | ({\rm fields}) S_\lambda(a,b) |0\rangle
  = n^{-\frac 14} \, 2^{\frac{1}{12}} \, e^{3 \zeta'(-1)} 
  \, (b{-}a)^{\frac18}
  \, \langle n | ({\rm fields}) \sigma_{+\lambda}(b)\,
  \sigma_{-\lambda}(a) |0\rangle
  ~\big( 1 + O(n^{-1}) \big)\,,
\labl{eq:S-twist-1/n}
where $\zeta$ is the Riemann zeta-function.
This shows that our formula for $S_\lambda(a,b)$ also has 
the correct large $n$ limit. 

In passing from $I_\lambda(a,b)$ to $S_\lambda(a,b)$ we have 
effectively decomposed the monodromy across the interval $[a,b]$
into a product of three terms, one each associated with one of
the lines in \erf{eq:contour-lens} (the explicit product can
be found in \erf{eq:tildeM} below). 
This is analogous to a method introduced in \cite{Deift1} to
analyse Riemann-Hilbert problems, which can also be applied
to investigate the asymptotics of orthogonal polynomials 
(see \cite{Bleher,Deift2}, and the lecture notes \cite{Deift3}). 
There, the orthogonal polynomials
are encoded in a Riemann-Hilbert problem with an appropriate
jump matrix across the real line, and their large-$n$
behaviour can be found by manipulating the contour along
which the jump condition is imposed in a way
analogous to \erf{eq:contour-lens}.

\medskip

The paper is organised as follows. In section 
\ref{sec:fields-int} we review how free boson vertex
operators and twist fields can be expressed as exponentials 
of integrated currents. To compare the properties of 
$I_\lambda(a,b)$ and twist fields, we calculate some
correlators of $su(2)_1$-currents in the presence of two twist
fields (section \ref{sec:twist}) and in the presence of $I_\lambda(a,b)$
(section \ref{sec:I_lam}). Finally, the definition of $S_\lambda(a,b)$ 
is given in section \ref{sec:Slam}, where also its properties are
investigated. Section \ref{sec:Out} contains our conclusions. We have 
also included two appendices where some of the more technical
calculations are described.

\sect{Representing fields as exponentiated integrals}\label{sec:fields-int}

Let us begin by explaining how one can represent fields in terms of
exponentiated integrals. While this may seem unfamiliar at first,
there is at least one example where this construction is actually well
known. This is the case of a free boson that we shall review first.

\subsection{The case of the $u(1)$ representation}

The free boson theory with field $X(z,\bar{z})$ has an $u(1)$ symmetry
that is generated by a weight one current 
$H(z)=i \, \partial X(z,\bar{z})$ with 
operator product expansion 
\be
H(z) \, H(w) = \frac{1}{2 (z-w)^2} + O(1) \,.
\ee
In terms of modes, $H(z)$ can be expanded as 
$H(z) = \sum_n H_n z^{-n-1}$. These modes then satisfy the commutation
relations  
\be
[H_m, H_n] = \frac{1}{2}\, m \, \delta_{m,-n} \,.
\ee
The corresponding stress energy tensor is 
$T(z) = \,{:}H(z) H(z){:}\,$, where the colons denote normal
ordering, {\it i.e.}
\be
{:}\,H(z) H(z)\,{:} ~= \lim_{w\rightarrow z} \, 
\Big(H(w) H(z) - \frac{1}{2 (w-z)^2} \Big) \,.
\labl{eq:u1-stresstens}
The modes of the stress energy tensor $T(z) = \sum_n L_n z^{-n-2}$ 
define a Virasoro algebra with $c=1$; in terms of the modes $H_n$ we
have 
\be
L_n = \sum_m : H_m \, H_{n-m} : \,,
\ee
where the colons denote here the usual normal ordering of modes. 
\smallskip

An (untwisted) highest weight representations of the $u(1)$
theory is generated by a state $|\mu\rangle$ that is annihilated
by the modes $H_n$ with $n>0$, and is an eigenvector of $H_0$ 
with eigenvalue $\mu$,
\be\label{hwr}
H_n\, |\mu\rangle = 
\mu \, \delta_{n,0}\, |\mu\rangle \,, \qquad n\geq 0 \,.
\ee
The corresponding vertex operator will be denoted by $V_\mu$, and
can be described by the usual vertex operator construction 
\be
V_\mu(z)  = \,{:}\,e^{2i\mu X(z)}\,{:}\,. 
\ee
We would now like to express this operator in terms of the current
$H(z)$ of the conformal field theory. At least formally we can write 
$i X(z) = \int^z H(w) dw$, and thus we should be able to write the 
vertex operator $V_\mu$ in terms of an exponentiated
integral. However, the exponentiated integral will have a 
non-vanishing vacuum expectation value, and thus it will not just 
describe the field $V_{\mu}$, but rather the pair of
$V_{\mu}$ together with its conjugate $V_{-\mu}$. Thus 
one is led to expect \cite{Gaberdiel:1998fs}  
\be\label{ansatz}
V_{\mu}(b) \,\, V_{-\mu}(a) = 
(b-a)^{-2 \mu^2} \,  
: \exp\!\Big( 2 \mu \int_a^b H(z) \, dz \Big) : \,,
\ee
where the prefactor is needed to produce the correct scaling
behaviour, as will be discussed further below (see (\ref{scaling})).
In fact, one can show that this identity holds in arbitrary
correlation functions. (For a definition of conformal field theory in
terms of correlation functions see for example \cite{Gaberdiel:1998fs}.) 
To this end one observes that the $V_{\pm \mu}$ satisfy indeed their 
defining relations \erf{hwr} since one calculates, using Wick's
Theorem, 
\bea
\Big\langle \prod_{i=1}^{n} H(u_i) \, H(w)\, 
: \exp\!\Big( 2 \mu \int_a^b \, H(z)\, dz \Big) :
\Big\rangle \enl
=~
\mu \, \Big( \frac{1}{w-b} - \frac{1}{w-a} \Big) \,
\Big\langle \prod_{i=1}^{n} H(u_i) \,
: \exp\!\Big( 2 \mu \int_a^b \, H(z)\, dz \Big) :
\Big\rangle \enl
\hspace*{3em} +~
\frac{1}{2} \sum_{j=1}^{n} \, \frac{1}{(w-u_j)^2} \, 
\Big\langle \prod_{i\ne j} H(u_i) \,
: \exp\!\Big( 2 \mu \int_a^b \, H(z)\, dz \Big) :
\Big\rangle \,,
\eear\ee
where we have used that 
\be
\int_a^b \frac{dz}{(w-z)^2} = \frac{1}{w-b} -  \frac{1}{w-a}  \,.
\ee
Since the $u_i$ are arbitrary, it follows that any correlation
function of $H(w)$ with the integrated exponential has only poles in
$(w-a)$ and $(w-b)$ of order one; the fields at $a$ and $b$ are
therefore highest weight states. It is also manifest from the above 
formula (by taking the contour integral around $a$ or $b$) that their
eigenvalues with respect to $H_0$ are $\pm \mu$. 

The above analysis determines the exponential up to an overall
function of $b-a$. This is fixed by considering the vacuum
expectation value of (\ref{ansatz}), which equals 
\be
\langle V_{\mu}(b) \, V_{-\mu}(a) 
\rangle = (b-a)^{-2 \mu^2} \,.
\labl{scaling}
This is of the form $(b-a)^{-2h}$ (as required by conformal symmetry)  
precisely for the choice of prefactor made in (\ref{ansatz}).
It is easy to see, by the same methods as above and using the
definition of the stress energy tensor in terms of the $u(1)$ field, 
that the fields that are defined by the right hand side of
(\ref{ansatz}) also have the correct conformal weight.
\smallskip

As a non-trivial consistency check, one can confirm that these highest
weight fields then also give rise to the correct $4$-point
functions. To this end we consider 
\bea
\big\langle V_{\mu}(b_1) \,\, V_{-\mu}(a_1) \,\,\,
V_{\nu}(b_2) \,\, V_{-\nu}(a_2) \big\rangle \enl
 = (b_1-a_1)^{-2\mu^2} \, (b_2-a_2)^{-2\nu^2} \,
\Big\langle 
{:}  \exp\!\Big( 2\mu \int_{a_1}^{b_1} \!\! H(z)\, dz \Big) {:} ~ 
{:}  \exp\!\Big( 2\nu \int_{a_2}^{b_2} \!\! H(w)\, dw \Big) {:} 
\Big\rangle \,. 
\eear\labl{1}
The correlator on the right hand side can now be easily evaluated and
one obtains 
\be\begin{array}{ll}\displaystyle
(\ref{1}) \!\!\! \etb  =  
(b_1-a_1)^{-2\mu^2} \, (b_2-a_2)^{-2\nu^2} \,
\sum_{l=0}^{\infty} \frac{2^l \mu^l \, \nu^l} {l!\, l!} \, 
\sum_{\sigma\in S_l}\, 
\Bigg(
\prod_{i=1}^{l} \int_{a_1}^{b_1} \!\!\! dz_i \, 
\int_{a_2}^{b_2} \!\!\! dw_i \Bigg) \, 
\prod_{j=1}^{l} \frac{1}{(z_j - w_{\sigma(j)})^2}
\enl
\etb = 
(b_1-a_1)^{-2\mu^2} \, (b_2-a_2)^{-2\nu^2} \,
\sum_{l=0}^{\infty} \frac{2^l \mu^l \, \nu^l} {l!}\, 
\prod_{i=1}^{l} \int_{a_1}^{b_1}\, dz_i
\left( \frac{1}{z_i-b_2} - \frac{1}{z_i-a_2} \right) \enl
\etb =  
(b_1-a_1)^{-2\mu^2} \, (b_2-a_2)^{-2\nu^2} \,
\left( \frac{ (b_1-b_2) \, (a_1-a_2)}{(a_1-b_2)\, (b_1-a_2)}
\right)^{2 \mu \nu}  \,.
\eear\ee
This then agrees with the known answer.

\subsection{Representations of $su(2)_1$}

At the self-dual radius the free boson theory is actually equivalent
to an $su(2)$ current theory with $k=1$. The $su(2)$ current symmetry
is generated by three currents $J^\pm$ and $J^3$ of conformal weight
one with operator product expansion\footnote{For $k=1$ this then
agrees with the above $u(1)$ theory by setting $J^3 = H$ and  
$J^\pm = V_{\pm 1}$.} 
\begin{eqnarray}
J^+(z) J^-(w) & = & \frac{k}{(z-w)^2} + \frac{2 J^3(w)}{z-w} + 
O(1) \nonumber \\
J^3(z) J^\pm(w) & = & \pm \frac{J^\pm(w)}{z-w} + 
O(1)  \\
J^3(z) J^3(w) & = & \frac{k}{2 (z-w)^2} + O(1)\,. \nonumber 
\end{eqnarray}
The modes of $J^3$ and $J^\pm$ then satisfy the commutation relations 
\begin{eqnarray}
{}[J^+_m,J^-_n] & = & 2 \, J^3_{m+n} 
+ k\, m\, \delta_{m,-n}  \nonumber \\
{}[J^3_m,J^\pm_n] & = & \pm\, J^\pm_{m+n}  \label{su2comm} \\
{}[J^3_m,J^3_n] & = & \frac{k}{2}\, m\, \delta_{m,-n} \,. \nonumber
\end{eqnarray}
At level $k=1$, the $su(2)$ theory has only two irreducible 
representations: the vacuum representation, and the
representation with $j=\tfrac12$. Here $j$ denotes the spin of the
$su(2)_k$ highest weight representation whose states of lowest
conformal weight are labelled by 
$|j,m\rangle$ with  $m=-j,-j+1, \ldots, j-1,j$. For $j=\tfrac12$ there
are only two highest weight states, namely $|\tfrac12,\tfrac12\rangle$ and  
$|\tfrac12,-\tfrac12\rangle$ that are conjugate to one another. It is
easy to see that the corresponding vertex operators $V(|j,m\rangle,z)$
can then be obtained by the previous construction provided we take 
$\mu=\pm \tfrac12$
\begin{equation}
V\left(|\tfrac12,\tfrac12\rangle,b\right)\, 
V\left(|\tfrac12,-\tfrac12\rangle,a\right) 
= (b-a)^{-\frac{1}{2}} \,  
: \exp\!\Big( \int_a^b \!\! J^3(z) \, dz \Big) : \,.
\end{equation}
For larger values of $k\geq 2$, however, the above construction does
not account for all non-trivial representations. In fact, in addition
to the vacuum representation (that corresponds to $\mu=0$) only
the highest weight state
$|\frac{k}{2},\frac{k}{2}\rangle$ together with its conjugate state 
$|\frac{k}{2},-\frac{k}{2}\rangle$ can be
described in the above manner (with $\mu=\tfrac12$).

\subsection{Twisted representations of the $su(2)_1$
  theory}\label{sec:twistrep} 

Up to now we have only discussed the untwisted highest weight
representations of $su(2)$. The corresponding vertex operators have
the property that the currents $J^3$ and $J^\pm$ are 
{\it single-valued} around the insertion point of the vertex
operator. As is well known, the $su(2)$ theory (like any affine
theory) has also {\it twisted representations} for which the currents
have non-trivial monodromies around the insertion points of the vertex
operators. For $su(2)$ all of these twisted representations are
actually equivalent (as affine representations) to untwisted
representation, since all automorphisms of SU(2) are inner. However
since the relevant identification modifies the energy momentum tensor,
twisted representations describe often different physical systems (for
an introduction to these matters see for example \cite{Goddard:1986bp}).  

In the following we shall mainly be interested in $\Zb_2$-twisted
representations of $su(2)_1$. One class of $\Zb_2$-twisted
representations have the property that the monodromy of the currents
is described by the (inner) automorphism
\begin{equation}\label{mon1}
J^3 \mapsto J^3\,, \qquad J^\pm \mapsto - J^\pm \,.
\end{equation}
The corresponding representation then has  modes $J^3_n$ that are
integer valued  $(n\in\Zb)$, while the modes  $J^\pm_r$ are
half-integer valued $(r\in\Zb+\frac{1}{2})$; these modes then still
satisfy the same commutation relations (\ref{su2comm}) as above. Since
these twisted representations are in one-to-one correspondence with the
usual untwisted representations, there are two inequivalent 
irreducible $\Zb_2$-twisted representations for $su(2)_1$:
they are generated from highest weight states $\tilde\sigma_\pm$ with   
$J^3_0 \tilde\sigma_\pm = \pm \frac{1}{4} \tilde\sigma_\pm$. From what 
was explained above, it is then clear that these vertex operators
can also be described by exponentiated integrals; indeed we have
simply 
\be
\tilde\sigma_+(b)\,\, \tilde\sigma_-(a) = (b-a)^{-\frac{1}{8}} \, 
: \exp\!\Big(\,\frac{1}{2} \int_a^b \!\! J^3(z)\, dz \,\Big) : \,.
\ee

In the following another class of $\Zb_2$-twisted representations will
play an important role: these are the $\Zb_2$-twisted representations
for which the monodromy is described by\footnote{see   
(\ref{eq:Z2-mono}) with $\lambda=1$. The case of 
general $\lambda$ differs from $\lambda=1$ by the rescaling
$J^\pm \mapsto \lambda^{\pm 1} J^\pm$. For the following it is
therefore sufficient to consider $\lambda=1$.}
\begin{equation}\label{mon2}
J^3 \mapsto - J^3\,, \qquad J^\pm \mapsto  J^\mp \,.
\end{equation}
Obviously, this only differs by a field redefinition from
(\ref{mon1}): if we define ({\it cf.}\ (\ref{eq:K-basis}) with
$\lambda=1$) 
\be
K^3  = \tfrac{1}{2} \big(J^+ + J^-\big) \,, \qquad
K^\pm = \pm \tfrac{1}{2} \big(J^+ - J^-\big) - J^3 \,,
\labl{eq:K1-basis}
then the fields $K^\pm$ and $K^3$ satisfy the same operator product
expansion as $J^\pm$ and $J^3$ (and thus their modes have the same
commutation relations as \erf{su2comm}). Furthermore, the monodromy
(\ref{mon2}) has the same form as (\ref{mon1}). In particular, the two
highest weight states  $\sigma_\pm$ that are now characterised by the
condition that $K^3_0\sigma_\pm = \pm \frac{1}{4} \sigma_\pm$ are
given by 
\be
\sigma_+(b)\,\, \sigma_-(a) ~=~ (b-a)^{-\frac{1}{8}} \, 
: \exp\!\Big(\,\frac{1}{4} \int_a^b \!\! 
     \big(J^+(z) + J^-(z)\big)\, dz \,\Big) :
\,. 
\labl{eq:twist-integral}
This is the formula that will motivate our ansatz for
$S_\lambda(a,b)$ (see section~5.1).

\sect{Properties of the $su(2)$-twist fields}\label{sec:twist}

As explained in the introduction, we propose that the operator
$I_\lambda(a,b)$ given in \erf{eq:Idef} can be expressed in terms of
twist fields as in \erf{eq:Sdef}. To support this claim, 
we shall compare, in section \ref{sec:Slam},
a number of properties of $I_\lambda(a,b)$ and $S_{\lambda}(a,b)$. As a
preparation,  we now want to study the correlation functions of the
$su(2)$ twist fields.

\subsection{Monodromy and zero-point function}\label{sec:mono-zero}

We will start with the zero-point function 
$\langle n|\sigma_+(b)\sigma_-(a)|0\rangle$. Here $\langle n|$ denotes
an out-state which is highest weight with respect to $J^3$ and has 
$J^3$-charge $n$,
\be
\langle n| \, J^3_m = n \, \delta_{m,0} \, \langle n| \qquad
\hbox{for $m\leq 0$.}
\ee
For $n \ge 0$ one can write $\langle n|$ explicitly in terms  
of $J^-$ modes (for $n \le 0$ one has to use $J^+$)
\be
  \langle n| = \langle 0| J^-_1 J^-_3 \cdots J^-_{2n-1} \,.
\labl{eq:n-via-J-}
Let us normalise $\langle 0|0 \rangle = 1$. 
One verifies that with the above definition $\langle n|n \rangle = 1$,
as well as
\be
  \langle n|J^-_m = 0 \quad {\rm for}~m \le 2n
  \qquad {\rm and} \qquad
  \langle n|J^+_m = 0 \quad {\rm for}~m \le -2n \,,
\ee
where we have used the null-vector relations of $su(2)_1$.
The highest weight property $\langle n|J^3_m = 0$ for $m<0$ and
$\langle n|J^3_0 = n \langle n|$ are then immediate consequences
of the commutation relations \erf{su2comm}. In order to evaluate
$\langle n|\sigma_+(b)\sigma_-(a)|0\rangle$ one can write 
the out-state in the form \erf{eq:n-via-J-} and express 
the $J^-$ modes in terms of the $K$-basis \erf{eq:K1-basis}.
For example, abbreviating 
$|\sigma\sigma\rangle \equiv \sigma_+(b)\sigma_-(a)|0\rangle$,
\be
  \langle 1 | \sigma\sigma\rangle
  = \langle 0| J^-_1 | \sigma\sigma\rangle
  = \langle 0| \big(K^3_1 - \tfrac{1}{2}(K^+_1 - K^-_1)\big) 
    | \sigma\sigma\rangle \,.
\ee
As is clear from the discussion in section \ref{sec:twistrep}, the field $K^3$
has only a simple pole with $\sigma_\pm$,  
\be
K^3(z) \, \sigma_{\pm}(a) = \pm \frac{1}{4} \frac{1}{(z-a)} + 
O(1) \,,
\ee
and thus it is easy to evaluate the term involving $K^3_1$, giving
$\langle 0| K^3_1 | \sigma\sigma\rangle = 
\tfrac14(b-a) \langle 0| \sigma\sigma\rangle$. For $K^\pm$ one
can write ({\it cf.}\ also \cite{Gaberdiel:1996kf})
\be
  \langle 0| K^\pm_1 | \sigma\sigma\rangle
  = \int_{\Cc_\infty} \mem z \, 
    \langle 0| K^\pm(z) | \sigma\sigma\rangle \, \frac{dz}{2\pi i}
  = \int_{\Cc_\infty} \mem \sqrt{(z-a)(z-b)} \,
    \langle 0| K^\pm(z) | \sigma\sigma\rangle \, \frac{dz}{2\pi i}
  = 0 \,.
\ee
To see the second equality, expand the square root in $z^{-1}$
and use the highest weight property of $\langle 0|$. In the third
step the contour $\Cc_\infty$ is deformed around the insertions
$\sigma_+(b)$ and $\sigma_-(a)$ and the highest weight property of
the latter is used. 

Altogether we thus obtain 
$\langle 1 | \sigma\sigma\rangle = 
\tfrac14(b-a) \langle 0| \sigma\sigma\rangle$. For higher values of
$n$ the calculation is similar, but more tedious. Luckily, the 
exact answer is known from factorising the correlator of four
twist fields \cite{Zam86,Zamolodchikov:1987ae},
\cite[eqn.\,(4.17)]{Dixon:1986qv},
\be
  \langle n|\sigma_+(b)\sigma_-(a)|0\rangle
  = 4^{-n^2} (b-a)^{n^2-\tfrac18} \,,
\labl{eq:n-twist}
where we normalised the twist fields such that
$\langle 0 | \sigma\sigma\rangle = 1 \cdot (b-a)^{-\tfrac18}$.

\subsection{Correlators involving $su(2)$-currents}

The next correlator we consider is the one-point function of
$J^3(z)$ in the presence of two twist fields. It is easily
determined by considering the function
\be
  f(z) = \sqrt{ (z-a)(z-b) } \,
  \langle n| \, J^3(z) \sigma_+(b) \sigma_-(a) \, |0\rangle
\ee
and noting that $f(z)$
is single valued on the complex plane and does not have any poles.  
Since $\langle n| J^3_0 = n \langle n|$ we furthermore have
$\lim_{z\rightarrow \infty} f(z) = 
 n \, \langle n| \, \sigma_+(b) \sigma_-(a) \, |0\rangle$
so that $f(z)$ is in fact a constant. In this way we find
\be
  \langle n| \, J^3(z) \sigma_+(b) \sigma_-(a) \, |0\rangle
  ~=~ \frac{n}{\sqrt{(z-a)(z-b)}} ~ 
  \langle n| \, \sigma_+(b) \sigma_-(a) \, |0\rangle \,.
\labl{eq:J3-2sig}
Correlators with several $J^3$ insertions can be determined
in a similar fashion.
\medskip

Finally we will need correlators with several $J^\pm$ insertions
in the presence of two twist fields. These can be determined
from the Knizhnik-Zamolodchikov equation, that is, by solving
the first order differential equations resulting from the
null-vector
\be
  (L_{-1} - 2 \nu J^3_{-1}) |J^\nu\rangle = 0 \,,
  \qquad {\rm where} \quad \nu \in \{\pm 1\}\,.
\ee
Knizhnik-Zamolodchikov equations in the presence of twist fields
have been studied in \cite{deBoer:2001nw}.
To find the differential equations first note that
as a consequence of this null-vector we have the identity
\be
  \int_{\Cc_z} \!\! \frac{\sqrt{(u{-}a)(u{-}b)}}{u-z} \, J^3(u)
  J^\nu(z) \, \frac{du}{2 \pi i}
  = 
  \frac{\nu}2 \Bigg( \frac{2z-a-b}{\sqrt{(z{-}a)(z{-}b)}}
  + \sqrt{(z{-}a)(z{-}b)} \, \frac{\partial}{\partial z} \Bigg) J^\nu(z) \,,
\labl{eq:cont-z}
where $\Cc_z$ is a contour winding closely around the point $z$. Here
we have used the operator product expansion 
\be
J^3(u) \, J^\nu(z) = \frac{\nu}{u-z} J^\nu(z) + 
(J^3_{-1}J^\nu)(z) + O(u-z) \,.
\ee
Around a point $w \neq z$ and around infinity we find analogously 
\bea
  \int_{\Cc_w} \!\! \frac{\sqrt{(u{-}a)(u{-}b)}}{u-z} \, J^3(u)
  J^\nu(w) \, \frac{du}{2 \pi i} ~=~ 
  \nu \, \frac{\sqrt{(w{-}a)(w{-}b)}}{w-z} \, J^\nu(w)
\enl
  \int_{\mathcal{C}_\infty} 
  \!\! \frac{\sqrt{(u{-}a)(u{-}b)}}{u-z} ~ \langle n | J^3(u)
  \,\frac{du}{2\pi i}
  ~=~ n \,\langle n | \,.
\eear\labl{eq:cont-inf}
Consider now the integral
\be
  \int_{\Cc_{z_k}} \!\! \frac{\sqrt{(u{-}a)(u{-}b)}}{u-z_k} \,
  \langle n | J^3(u) J^{\nu_1}(z_1)\cdots J^{\nu_m}(z_m)
  |\sigma\sigma\rangle \,
  \frac{du}{2 \pi i} \,.
\ee
This contour integral can be calculated in two ways: on the one hand, 
we can directly use \erf{eq:cont-z} and thus evaluate the contour
integral in terms of the right hand side of \erf{eq:cont-z}. On the
other hand, we can deform the contour around $z_k$ to encircle all
other insertion points  $z_i$, $i \neq k$ as well as infinity and $a$
and $b$. As in the previous calculation, there is no contribution from
the twist-field insertions at $a$ and $b$. The individual
contributions from the points $z_i$ and infinity can be evaluated
using \erf{eq:cont-inf}, and one thus arrives at the system of partial
differential equations 
\be
  D_k \, \langle n | J^{\nu_1}(z_1)\cdots J^{\nu_m}(z_m)
  |\sigma\sigma\rangle = 0 \qquad \hbox{for $k =1,\dots,m$,}
\labl{eq:KZ-twist}
with
\be
  D_k = \frac{\nu_k}2 \Bigg( \frac{2z_k-a-b}{\sqrt{(z_k{-}a)(z_k{-}b)}}
  + \sqrt{(z_k{-}a)(z_k{-}b)} \, \frac{\partial}{\partial z_k} \Bigg)
  - n + \sum_{i \neq k} 
  \nu_i \frac{\sqrt{(z_i{-}a)(z_i{-}b)}}{z_i-z_k} \,.
\labl{eq:KZ-twist-D}
To solve the equations \erf{eq:KZ-twist} it is convenient to
pass from the complex plane with twist-field insertions at $a$
and $b$, which generate a branch cut on the interval $[a,b]$,
to a double cover ({\it cf.}\ 
\cite{Dixon:1986qv,Zamolodchikov:1987ae}).  
The letters $z$, $w$ will always refer to points on the 
complex plane which forms the base of the double cover and
$\zeta$, $\xi$ to points on the double cover. Since we have a
single cut, the double cover has again the 
topology of a Riemann sphere and we fix the projection from the double
cover to the base by 
\be
  z = \frac{b-a}4 \big( \zeta + \zeta^{-1} \big) + \frac{a+b}2 \,.
\ee
Conversely, the two pre-images of a point $z \notin [a,b]$ are given
by $\zeta(z)$ and $\zeta(z)^{-1}$ with
\be
  \zeta(z) = \frac{ \big( \sqrt{z{-}a} + \sqrt{z{-}b} \big)^2 }{b-a} \,.
\labl{eq:zeta(z)}
For all square roots we choose the convention that there is
a branch cut from $-\infty$ to $0$ and that $\sqrt{1}=1$. 
Then $z \neq [a,b]$ implies $|\zeta(z)|>1$. In fact, if we
write $z = x + iy + \tfrac{a+b}2$, 
the curve $|\zeta(z)|=r>1$ is the following ellipse with centre 
$\tfrac{a+b}2$,
\be
  \frac{x^2}{r_x^{\,2}} +
  \frac{y^2}{r_y^{\,2}} = 1 \,,
 \qquad 
  r_x = \frac{b{-}a}{2} ~ \frac{r^2{+}1}{2r} \,,\quad
  r_y = \frac{b{-}a}{2} ~ \frac{r^2{-}1}{2r} \,.
\labl{eq:modzeta-ellipse}
This ellipse can also be described by $|z{-}a|+|z{-}b|=
\tfrac{b{-}a}2\big( r+ r^{-1} \big)$, which shows that $a$ and
$b$ are the two focal points of the ellipse \erf{eq:modzeta-ellipse}.
The significance of this parametrisation is that
the contour integrals defining \erf{eq:Idef} will be  
suppressed by factors $r^{-2n}$ when carried out 
along the ellipse (see section \ref{sec:non-pert}).  

On the double cover, each $J^-$-field corresponds
to the pair of vertex operators 
$V_{-}(\zeta)\, V_{+}(\zeta^{-1})$, where  
$V_\pm(\zeta)$ denotes the vertex operator
$V_{\pm 1/\sqrt{2}}(\zeta)$ of $J^3$-charge 
$\pm \tfrac{1}{\sqrt{2}}$. Thus one expects that 
\bea
  \frac{\langle n | J^{-}(z_1)\cdots J^{-}(z_m)
  |\sigma\sigma\rangle}{\langle n |\sigma\sigma\rangle} 
\enl
  \qquad = \Big(\frac{4}{b-a}\Big)^m
  \prod_{i=1}^m \frac{1}{\zeta_i-\zeta_i^{-1}}
  \,\big\langle \tfrac{n}{\sqrt{2}} \big| 
  V_-(\zeta_1) V_+(\zeta_1^{-1}) \cdots
  V_-(\zeta_m) V_+(\zeta_m^{-1}) 
  \big|\tfrac{n}{\sqrt{2}}\big\rangle \,,
\eear\labl{eq:J-corr-via-V}
where $\zeta_i=\zeta(z_i)$. The right hand side can be computed in
terms of the Coulomb gas expression for free boson vertex operators,
and one finds 
\be
\frac{ 
\langle n| J^-(z_1) \cdots J^-(z_m) 
\sigma^+(b) \sigma^-(a) |0\rangle}{
\langle n|\sigma^+(b) \sigma^-(a) |0\rangle}
=
\Big(\frac{4}{b-a}\Big)^{m}
\prod_{i=1}^m \frac{\zeta_i^{-2n}}{(\zeta_i-\zeta_i^{-1})^2}
\prod_{i>j} \Big(\frac{\zeta_i-\zeta_j}{\zeta_i \zeta_j-1}\Big)^2 \,.
\labl{eq:J-corr-twist}
It is straightforward to check that this indeed solves 
\erf{eq:KZ-twist}; to this end it is useful to rewrite the
differential operator \erf{eq:KZ-twist-D} also in terms of the
$\zeta$-variables, using 
$\zeta-\zeta^{-1} = \tfrac{4}{b-a}\sqrt{(z-a)(z-b)}$ and 
$\tfrac{\partial}{\partial \zeta} 
= \tfrac{b-a}{4}(1-\zeta^{-2}) \tfrac{\partial}{\partial z}$.
Of course, \erf{eq:KZ-twist} only determines 
\erf{eq:J-corr-twist} up to a constant. This constant can be
found recursively using \erf{eq:n-via-J-},
\be
\langle n{+}1| J^-(z_1) \cdots J^-(z_m) |\sigma\sigma\rangle
=
\int_{\Cc_\infty} \mem w^{2n+1} \,
\langle n| J^-(w) J^-(z_1) \cdots J^-(z_m) |\sigma\sigma\rangle 
\,\frac{dw}{2\pi i} \,.
\ee
This determines the overall constant for a correlator with 
$m{+}1$ insertions of $J^-$ in terms of a correlator with only
$m$ insertions. Finally, the expressions with no insertions
of $J^-$ has already been given in \erf{eq:n-twist}.
This procedure is the origin of the factor 
$\big(\tfrac{4}{b-a}\big)^m$ in \erf{eq:J-corr-via-V} and
\erf{eq:J-corr-twist}.

A correlator where some of the $J^-(z_i)$ insertions have been
replaced by $J^+(z_i)$ insertions can be obtained by continuing
$z_i$ through the branch cut $[a,b]$, which amounts to replacing
$\zeta_i \rightarrow \zeta_i^{-1}$ in \erf{eq:J-corr-twist}.

\sect{Properties of the operator $I_\lambda(a,b)$}\label{sec:I_lam}

In the previous section we have collected some information about the
structure of the twist field correlators. Now we want to study the 
correlation functions of $I_\lambda(a,b)$ defined in \erf{eq:Idef}.

\subsection{Monodromy and operator product expansion of
  $su(2)$-currents}\label{sec:I-mono-OPE} 

Before computing the monodromy of the $su(2)$-currents
$J^\pm$, $J^3$ in the presence of $I_\lambda(a,b)$ let
us consider a slightly more general situation. Define the
operator 
\be
  B(a,b) = \exp\!\Big( 
  \int_a^b \!\!F(x) \, dx \Big) \,,
\labl{eq:Bdef}
where $F(x)$ is a linear combination of holomorphic functions 
multiplying chiral fields (and we assume that the operator product
expansion of $F(x_1)$ and $F(x_2)$ does not have any poles so that no
normal ordering prescription is necessary). Consider the analytic
continuation of a chiral field $\phi(z)$ around the point $b$, 
\be
  \begin{picture}(400,50)(0,0)
  \put(0,0){
    \put(0,0){\scalebox{0.35}{\includegraphics{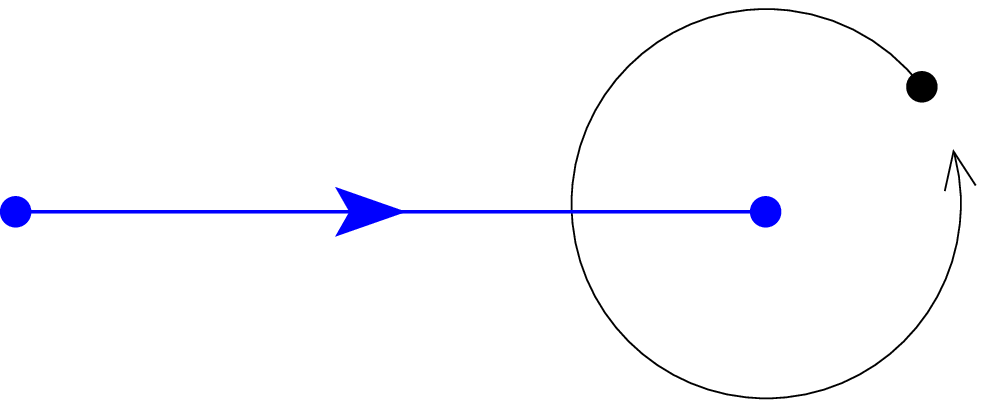}}}
    \put(0,9){$a$}
    \put(75,9){$b$}
    \put(98,30){$\phi(z)$}
  }
  \put(140,19){
    \put(0,0){\vector(1,0){70}}
    \put(10,3){anal.\,cont.}
    }
  \put(250,0){
    \put(0,17){\scalebox{0.35}{\includegraphics{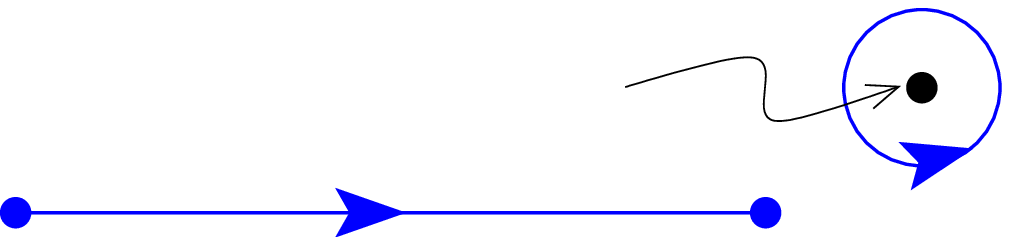}}}
    \put(0,9){$a$}
    \put(75,9){$b$}
    \put(43,29){$\phi(z)$}
    \put(100,20){$\Cc_z$}
  }
  \end{picture}
\labl{eq:cont-around-b}
It is then clear that under the analytic continuation
\erf{eq:cont-around-b}, the monodromy of $\phi(z)$ is 
\be
  \phi(z) \longrightarrow 
  \exp\!\Big( 2 \pi i 
  \int_{\Cc_z} \!\!F(x) \,\frac{dx}{2\pi i} \Big) \,\phi(z) \,. 
\labl{eq:generic-mono}
In the case of $I_\lambda(a,b)$ we have 
$F(x) = \tfrac{\lambda}{2\pi} J^+(x)$. Acting on the state
$|\phi\rangle$ corresponding to the field $\phi(z)$,
the monodromy \erf{eq:generic-mono} then reads
$|\phi\rangle \rightarrow \exp( i \lambda J^+_0)|\phi\rangle$.
Representing the linear combination
$|\phi\rangle = \alpha |J^+\rangle + \beta |J^3\rangle + 
\gamma |J^-\rangle$ by the vector $(\alpha,\beta,\gamma)$,
the monodromy of the $su(2)$-currents around the endpoint $b$
of $I_\lambda(a,b)$ is given by the matrix
\be
  M_\lambda(b) = 
 \begin{pmatrix} 1 & -i \lambda & \lambda^2 \\ 
                 0 & 1 & 2i \lambda \\ 0 & 0 & 1 
 \end{pmatrix} \,~.
\labl{eq:I-mono}
For example, 
$\exp( i \lambda J^+_0)|J^3\rangle
= (J^3_{-1} + i \lambda J^+_0 J^3_{-1})|0\rangle
= |J^3\rangle - i \lambda |J^+\rangle$.
Since the currents are single valued on $\Cb - [a,b]$,
the monodromy around $a$ is inverse to that around $b$
so that $M_\lambda(a) = M_\lambda(b)^{-1} = M_{-\lambda}(b)$.

The above calculation can be repeated for the stress tensor $T(z)$
and one finds that due to $[J^+_0,L_m] = 0$, the stress tensor
is single valued across $[a,b]$.
Translating this observation back into matrix model language
(as briefly described in section \ref{sec:mat-2dgrav}), gives
a way to derive the quadratic loop equation of the matrix model
from the free boson conformal field theory
\cite{Kharchev:1992iv, Morozov:hh, Kostov:1999xi}. 

\medskip

To analyse the singularities of the $su(2)$-currents close to
the points $a$ and $b$ it is helpful to introduce the following
combinations of fields
\be\bearll\displaystyle
 \widehat J^+(z) \etb=~ J^+(z) 
 \enl
 \widehat J^3(z) \etb=~ J^3(z) + \tfrac{\lambda}{2\pi} 
   \ln\tfrac{z{-}b}{z{-}a} \, J^+(z) 
 \enl
 \widehat J^-(z) \etb=~ J^-(z) - \tfrac{2\lambda}{2\pi}  
   \ln \tfrac{z{-}b}{z{-}a} \, J^3(z) 
 - \Big(\tfrac{\lambda}{2\pi} \ln \tfrac{z{-}b}{z{-}a}\Big)^2 J^+(z) \,.
\eear\labl{eq:hat-J}
Using \erf{eq:I-mono} one verifies that $\widehat J^\pm(z)$ and
$\widehat J^3(z)$ are single valued in the presence of a single
insertion of $I_\lambda(a,b)$. In particular, they can be expanded
in integer modes
\be
  \widehat J^{c}(z) = \sum_{m \in \Zb} (z-p)^{-m-1} \widehat
  J^{c}_{m;p} \,,   \qquad {\rm where} \qquad
  \widehat J^{c}_{m;p} = \int (z-p)^m 
\widehat J^{c}(z) \frac{dz}{2\pi i} \,.
\ee
Here $c \in \{+,3,-\}$ and $p$ equals $a$ or $b$.
We would like to establish the following statement:
\begin{itemize}
\item[(S)] Suppose $\lim_{z\rightarrow b} (z-b) J^-(z) I_\lambda(a,b)$  
is finite inside any correlator, and 
there exists some $N > 0$ such that
$\lim_{z\rightarrow b} (z-b)^N J^+(z) I_\lambda(a,b)$
and 
$\lim_{z\rightarrow b} (z-b)^N J^3(z) I_\lambda(a,b)$
are zero.
Then
\be
  \widehat J^{\pm}_{m;b} I_\lambda(a,b) = 0 
  = \widehat J^{3}_{m;b}I_\lambda(a,b) \quad
    \hbox{for $m>0$}\,, \qquad
  \widehat J^{+}_{0;b} I_\lambda(a,b) = 0
  = \widehat J^{3}_{0;b} I_\lambda(a,b) \,.
\labl{eq:I-hw}
\end{itemize}
Together with the analogous statement for $a$,
this means that the fields at the end points $a$ and $b$ are 
in fact highest weight with respect to the $\widehat{J}^c$ action. 

The vanishing of the $J^+$-zero mode implies 
in particular that 
$J^+(z) I_\lambda(a,b)$ is regular for $z \rightarrow b$.
Of course, this also follows immediately when writing out
the integrals in \erf{eq:Idef} and using that the operator product
expansion of $J^+$ with itself is regular. However, in section
\ref{sec:S-pole} we will need to apply statement (S) to
$S_\lambda(a,b)$ instead of $I_\lambda(a,b)$, so we will
present the argument in a form which will be valid also then.

To establish (S), start by expressing $J^-$ in terms
of the single valued combinations \erf{eq:hat-J},
\be
  J^-(z) = \widehat J^-(z) + 
  \tfrac{2\lambda}{2\pi}  \ln \tfrac{z{-}b}{z{-}a} \, \widehat J^3(z) 
  - \Big(\tfrac{\lambda}{2\pi} \ln \tfrac{z{-}b}{z{-}a}\Big)^2 
  \widehat J^+(z)\,.
\labl{eq:J-viaJhat}
Denote by $C(\phi)$ a correlator involving $\phi$ and 
$I_\lambda(a,b)$, as well as any number of other fields. Expanding the
right hand side of \erf{eq:J-viaJhat} in terms of modes around $b$
then gives 
\be
  C(J^-(z)) = \sum_{m = -\infty}^N (z-b)^{-m-1} 
  \Big( C(\widehat J^-_{m;b}) +
  \tfrac{2\lambda}{2\pi}  \ln \tfrac{z{-}b}{z{-}a} \, 
  C(\widehat J^3_{m;b}) 
  - \Big(\tfrac{\lambda}{2\pi} \ln \tfrac{z{-}b}{z{-}a}\Big)^2 
  C(\widehat J^+_{m;b}) \Big)~.
\ee
The summation is truncated by the assumption on the limit
$z\rightarrow b$ of $J^+$ and $J^3$. Evaluating the conditions 
$\lim_{z\rightarrow b} (z-b)^{m+1} J^-(z) I_\lambda(a,b) = 0$  
for $m=N,N{-}1,\dots,1$ gives $C(\widehat J^{\pm,3}_{m;b})=0$ 
for that range of $m$. 
Finally, for $m=0$ we get 
$C(\widehat J^+_{0;b})=0=C(\widehat J^3_{0;b})$.
Since $C(\dots)$ was an arbitrary correlator, this implies
\erf{eq:I-hw}.
\smallskip

In order to establish the highest weight relations \erf{eq:I-hw}
for $I_\lambda(a,b)$ we still have to verify that the conditions
of the statement (S) are met. For $J^+$ this is obvious, but for $J^3$
and $J^-$ this requires a short calculation which is given
in appendix \ref{sec:I-pole-cond}.

\medskip

Finally, let us show that the endpoints of $I_\lambda(a,b)$ obey
the Virasoro highest weight condition for weight zero. To this end,
instead of expressing the stress tensor $T(z)$ 
as in \erf{eq:u1-stresstens} we
use the single valued fields \erf{eq:hat-J}. Computing the operator
product expansion of $\widehat J^3$ with itself to order $O(z{-}w)$
one finds 
\be
  T(w) = \lim_{z \rightarrow w} 
  \Big( \widehat J^3(z) \widehat J^3(w) - \frac{1/2}{(z{-}w)^2}\Big)
  + \frac{\lambda}{2\pi} 
     \frac{b-a}{(w{-}a)(w{-}b)} \widehat J^+(w) \,,
\ee
which in terms of modes around $b$ reads
\be
  L_{m;b} = \sum_{k\in\Zb} : \widehat J^3_{k;b} \widehat J^3_{m-k;b} :
  + \frac{\lambda}{2\pi} \sum_{k=0}^\infty (a{-}b)^{-k} 
  \widehat J^+_{m+k;b} \,.
\ee
Together with \erf{eq:I-hw} this immediately implies that
$L_{m;b} I_\lambda(a,b) = 0$ for $m \ge 0$. For $a$ instead of $b$
one finds the same result.

\subsection{Zero-point function}\label{sec:I-zero-pt}

Up to now we have kept the parameter $\lambda$ arbitrary. But
just as was the case for twist fields, we can restrict our
attention to the case $\lambda = 1$. The results for general
values of $\lambda$ are then obtained via the identity
\be
  I_\lambda(a,b) |0\rangle = e^{ \ln(\lambda) J^3_0 }
  I_1(a,b) |0\rangle \,.
\labl{eq:Ilam-I1}
To see this write $I_\lambda(a,b)$ and $I_1(a,b)$ as a sum
over $J^+$ integrations and use $[J^3_0,J^+(z)] = J^+(z)$ which
leads to $\exp( \ln(\lambda) J^3_0 ) J^+(z) = 
\lambda  J^+(z) \exp( \ln(\lambda) J^3_0 )$. Moving
the exponential in \erf{eq:Ilam-I1} past the $J^+$ insertions
in the expansion of $I_1(a,b)$ gives a factor of $\lambda$ for
each insertion.

\medskip

The correlator we would like to compute is
$\langle n| I_1(a,b) |0\rangle$. From the previous section we
know that the endpoints of $I_1(a,b)$ behave as Virasoro primary
fields of weight zero. Applying a rescaling and a translation yields
\be
  \langle n| I_1(a,b) |0\rangle = (b{-}a)^{n^2} 2^{-n^2}
  \langle n| I_1(-1,1) |0\rangle \,.
\ee
The correlator on the right hand side can be computed by explicit 
integration using orthogonal polynomials on the interval $[-1,1]$,
{\it i.e.}\ the Legendre polynomials (see 
{\it e.g.}\ \cite{DiFrancesco:1993nw,Fyodorov:2004ar}
for an introduction to the method
of orthogonal polynomials), or by directly using Selberg's integral
(see {\it e.g.}\ \cite[chapter 17]{Mehta}). The result is
\be
\begin{array}{rcl}
\langle n| I_1(-1,1) |0\rangle 
& = & {\displaystyle \frac1{(2\pi)^{n}\,n!}
  \int_{-1}^1 \!\!\!\! dx_1\cdots dx_n ~
  \langle n| J^+(x_1)\cdots J^+(x_n) |0\rangle } \vspace{0.2cm}\\
& = & {\displaystyle \frac1{(2\pi)^{n}\,n!}
  \int_{-1}^1 \!\!\!\! dx_1\cdots dx_n  \, \Delta(x)^2
  = (2\pi)^{-n} \, 2^{n^2} \det(H_n) \,.}
\end{array}
\ee
In the first step, the definition \erf{eq:Idef} has been substituted,
in the second step the Coulomb gas expression for the integrand has
been written in terms of the Vandermonde determinant
$\Delta(x) = \prod_{i>j} (x_i{-}x_j)$. In the result, $H_n$ is the 
$n{\times}n$-Hilbert matrix $(H_n)_{ij} = (i+j-1)^{-1}$. Its determinant
is given by
\be
  \det(H_n) 
  = 2^{-n^2} \prod_{k=0}^{n-1} \frac{2}{2k{+}1}
  \Bigg(\frac{2^k \, k!^2}{(2k)!} \Bigg)^{\!2} 
  = \frac{2^{n-2n^2} \pi^{n+1/2}}{\Gamma(n{+}\tfrac12)}
  \Bigg( \frac{G(n{+}1)\,G(\tfrac12)}{G(n{+}\tfrac12)} \Bigg)^{\!2} \,.
\ee
Here $G(z)$ is the Barnes function, which is defined by
$G(z{+}1) = \Gamma(z) G(z)$ and $G(1)=1$  
together with a convexity condition. The behaviour of 
$\det(H_n)$ for large $n$ can now be obtained from the large $z$
expansion of $G(z)$.
The latter can be found in \cite[eqn.\,(28)]{Adamchik}
(in the preprint (v1), the minus in front of the sum should be 
a plus) or in \cite[eqn.\ (2.38)]{Marino:2004eq}. 
In this way, we finally
obtain
\be
  \langle n| I_1(a,b) |0\rangle = (b{-}a)^{n^2} 4^{-n^2}
  n^{-\frac14} \, 2^{\frac1{12}} \, e^{3 \zeta'(-1)}
  \exp\!\Big(  
  - \tfrac{1}{64} n^{-2}
  + \tfrac{1}{256} n^{-4} + O(n^{-6}) \Big) \,.
\labl{eq:n-I1}
Comparing to the correlator of two twist fields \erf{eq:n-twist} we
observe that $\sigma_+(b)\sigma_-(a)$ does correctly reproduce the leading
term in \erf{eq:n-I1} in the sense that
\be
  \lim_{n \rightarrow \infty} n^{-2}
  \ln \langle n| \sigma_+(b)\sigma_-(a) |0\rangle
  =
  \lim_{n \rightarrow \infty} n^{-2}
  \ln \langle n| I_1(a,b) |0\rangle \,.
\ee

\subsection{One-point function}\label{sec:one-point}

After computing the zero-point function we will turn to the
correlators $\langle n| J^c(z) I_1(a,b) |0\rangle$ for
$c \in \{+,3,-\}$. The result can again be obtained by
matrix model techniques using orthogonal polynomials
(see {\it e.g.}\ \cite[chapter 22]{Mehta} or 
\cite[section 4]{Fyodorov:2004ar}), which amount to 
explicitly computing
the relevant integrals. Another method more in the spirit
of conformal field theory is to use that the $h=1$, $c=1$
Virasoro highest weight representation has a null vector
at level three. Denoting the highest weight vector in this
representation by $|J\rangle$ the null vector $|\eta\rangle$ is
\be
  |\eta\rangle = 
 \big( 3 L_{-3} - 2 L_{-1} L_{-2} + 
 \tfrac12 L_{-1}L_{-1}L_{-1} \big) |J\rangle = 0 \,.
\labl{eq:level3-null}
Since the three $su(2)$ currents $J^c(z)$ are Virasoro
primaries of weight one, the three correlators
$\langle n| J^c(z) I_1(a,b) |0\rangle$ will all solve the
same third order differential equation in $z$, obtained from
the null vector $|\eta\rangle$. To compute this differential
equation, recall from section \ref{sec:mono-zero} that from the
point of view of the Virasoro algebra there is no difference between
an insertion of $I_\lambda(a,b)$ and a product $\phi(b)\phi(a)$ of
Virasoro primary fields $\phi$ with conformal weight zero.
Specialising to $a=-1$ and $b=1$,
the differential equation is then found to be
(see \cite[section 8.3]{DiFrancesco:nk} for more details on null-vector
computations)
\bea
  0 = \langle n| \, \eta(z) \, \phi(1) \, \phi(-1) \, |0\rangle
  \enl
  \phantom{0} = \Bigg\{ \frac12 \partial_z^{\,3}
  + \frac{4z}{z^2{-}1} \partial_z^{\,2}
  + \frac{5z^2{-}1}{(z^2{-}1)^2} \partial_z
  - \frac{2(n^2{-}1)}{z^2{-}1} \Big( \frac{z}{z^2{-}1} 
  + \partial_z \Big) \Bigg\} 
  \langle n| J(z) \phi(1) \phi(-1) |0\rangle \,.
\eear\labl{eq:null-deq}
The space of solutions to this equation is three-dimensional.
The elements in this space describing the three functions
$\langle n| J^c(z) I_1(-1,1) |0\rangle$ have to be identified
from their behaviour at the singular points $-1$, $1$, $\infty$.

In any case, using either orthogonal polynomials or Virasoro null
vectors, the final result for the one-point functions is
\be\bearll
  \langle n| J^+(z) I_1(-1,1) |0\rangle
  \etb=~ 
  \pi n  
  \Big(P_{n-1}(z) P_n'(z) - P_{n}(z) P_{n-1}'(z) \Big)  
  \langle n| I_1(-1,1)|0\rangle 
  \enl
  \langle n| J^3(z) I_1(-1,1) |0\rangle
  \etb=~ 
  n \Big( P_{n-1}(z) Q_{n}'(z) - P_{n}(z) Q_{n-1}'(z) \Big)  
  \langle n| I_1(-1,1)|0\rangle 
  \enl
  \langle n| J^-(z) I_1(-1,1) |0\rangle
  \etb=~ 
  -\frac{n}{\pi} \Big(Q_{n-1}(z) Q_n'(z) - Q_n(z) Q_{n-1}'(z) \Big) 
   \langle n| I_1(-1,1)|0\rangle \,.
\eear\labl{eq:n-J-I1}
In these equations, $P_n(z)$ is the $n$'th Legendre polynomial and
$Q_n(z)$ the $n$'th Legendre function of the second kind. The first
few are (chosen such that they are real for $z \in \Rb-[-1,1]$)
\be
  Q_0(z) = \frac12 \ln\!\Big(\,\frac{z{+}1}{z{-}1}\,\Big) \,, \qquad
  Q_1(z) = \frac{z}2 \ln\!\Big(\,\frac{z{+}1}{z{-}1}\,\Big)
   -1\,, \qquad
  Q_2(z) = \frac{3z^2{-}1}4 \ln\!\Big(\,\frac{z{+}1}{z{-}1}\,\Big)
   - \frac{3z}{2}  \,.
\ee
One can verify that the functions \erf{eq:n-J-I1} solve 
\erf{eq:null-deq} and their monodromy around the point $1$ 
is given by \erf{eq:I-mono}.

To understand the large-$n$ behaviour of the one-point functions,
it is convenient to write $P_n(z)$ and $Q_n(z)$ for $z \notin [-1,1]$
in terms of hypergeometric functions as \cite[eqn.\,(8.723)]{GrRy}
\bea
  P_n(z) = \frac{ \Gamma(n{+}\tfrac12) }{ \sqrt{\pi} \, \Gamma(n{+}1) }
  \, \frac{ \zeta^{n+1} }{\sqrt{\zeta^2{-}1}} \, 
  {}_2\!F\!{}_1\!\big( \tfrac12 , \tfrac 12 ; \tfrac12{-}n ;
  \tfrac{-1}{\zeta^2{-}1} \big) 
  \enl
  Q_n(z) = \frac{ \sqrt{\pi} \, \Gamma(n{+}1) }{ \Gamma(n{+}\tfrac32) }
  \, \frac{ \zeta^{-n} }{\sqrt{\zeta^2{-}1}} \, 
  {}_2\!F\!{}_1\!\big( \tfrac12 , \tfrac 12 ; \tfrac32{+}n ;
  \tfrac{-1}{\zeta^2{-}1} \big) 
  \,.
\eear\labl{eq:leg-hyper}
Here, $\zeta = \zeta(z)$ is given by \erf{eq:zeta(z)} with $a=-1$, $b=1$. 
The large-$n$ expansion of the correlators \erf{eq:n-J-I1} is then 
found by writing out the definition of the hypergeometric functions 
\erf{eq:leg-hyper} as a power series and expanding the Gamma-functions,
\be\bearll
  \frac{\langle n| J^+(z) I_1(-1,1) |0\rangle}{\langle n| I_1(-1,1)
    |0\rangle} 
  \etb \!\!=~ 
  \frac{2\,\zeta^{2n}}{(\zeta{-}\zeta^{-1})^{2}}
  \Big( 1 + \frac{1}{4} \, \frac{\zeta+\zeta^{-1}}{\zeta-\zeta^{-1}}
  \, n^{-1}   + \frac{1}{32} \,
  \frac{(1{+}5\zeta^2)(1{+}5\zeta^{-2})}{(\zeta-\zeta^{-1})^2} \,
  n^{-2} 
  + O(n^{-3}) \Big)
\enl
  \frac{\langle n| J^3(z) I_1(-1,1) |0\rangle}{\langle n| I_1(-1,1)
    |0\rangle} 
  &\!\!=~ \displaystyle
   \frac{2 n}{\zeta-\zeta^{-1}}\Big(
   1 - \frac{1}{2(\zeta{-}\zeta^{-1})^2} \, n^{-2} + O(n^{-3}) \Big) 
\enl
  \frac{\langle n| J^-(z) I_1(-1,1) |0\rangle}{\langle n| I_1(-1,1)
    |0\rangle} 
   &\!\!=~ \displaystyle
  \frac{2\,\zeta^{-2n}}{(\zeta{-}\zeta^{-1})^{2}}
  \Big( 1 - \frac{1}{4} \, \frac{\zeta+\zeta^{-1}}{\zeta-\zeta^{-1}}
  \, n^{-1} 
  + \frac{1}{32} \,
  \frac{(1{+}5\zeta^2)(1{+}5\zeta^{-2})}{(\zeta-\zeta^{-1})^2} \,
  n^{-2} 
  + O(n^{-3}) \Big) \,.
\eear\labl{eq:n-J-expand}
Comparing to \erf{eq:J3-2sig} and \erf{eq:J-corr-twist} 
(specialised to $a=-1$ and $b=1$) we see again that
the leading behaviour in $1/n$ is the same for 
$I_1(a,b)$ and the twist fields $\sigma_+(b)\sigma_-(a)$. 

In fact not only the leading term in the expansions
\erf{eq:n-J-expand} has the $\Zb_2$-symmetry $J^\pm \rightarrow J^\mp$
and $J^3 \rightarrow -J^3$ upon analytically continuing $z$ through
the branch cut $[-1,1]$, but this monodromy is retained at any finite
order in the expansions \erf{eq:n-J-expand}. 
It is only after summing all terms that the `correct' monodromy
\erf{eq:I-mono} is recovered. 
For example, at finite values of $n$, the correlator
$\langle n| J^+(z) I_1(-1,1) |0\rangle$ is a single valued function
of $z$ (in fact, a polynomial), while in the $1/n$-expansion
it has a branch cut on $[-1,1]$. 
Recall from \erf{eq:mm-well-1pt} that this conformal field theory
correlator is related to the correlator
$Z_{\rm mm}^{\rm well}[ {\rm det}(z-M)^{2} ]^{(n)}$ 
in the hermitian one-matrix model, and hence the latter shares with
the correlator of $J^+(z)$ the property that the monodromy of the
individual
terms in the $1/n$-expansion differs from the monodromy of the
complete expression. This is a manifestation of Stokes' phenomenon
as mentioned at the end of section \ref{sec:mat-2dgrav}.

\sect{Writing $I_\lambda(a,b)$ in terms of twist
  fields}\label{sec:Slam} 

In sections \ref{sec:twist} and \ref{sec:I_lam} we have collected 
some properties of $su(2)$-twist fields and of the operator
$I_\lambda(a,b)$. 
We have seen that in sectors of large $J^3$-charge, $I_\lambda(a,b)$
behaves very similar to a product $\sigma_+(b)\sigma_-(a)$ of twist
fields. Motivated by these observations one can now try to find an
alternative expression for the operator $I_\lambda(a,b)$ in terms of
appropriately dressed twist fields. 

To achieve this, we make an ansatz $S_\lambda(a,b)$ involving 
twist fields and $J^-$-integrals in section \ref{sec:I-mono}, and show
that it can reproduce the correct monodromy. That the $su(2)$-currents
have the same monodromy for $I_\lambda(a,b)$ and $S_\lambda(a,b)$ is
the first piece of evidence for the proposed identity
$I_\lambda(a,b)=S_\lambda(a,b)$. The $J^-$ integrals in 
$S_\lambda(a,b)$ are divergent and need to be regulated; this is done 
in section 
\ref{sec:regulate}. Next, in section \ref{sec:non-pert} 
we investigate the large-$n$ behaviour of $S_\lambda(a,b)$
and find that it is given in terms of a product of twist fields
$\sigma_+(b)\sigma_-(a)$. This is the second piece of
evidence for $I_\lambda(a,b)=S_\lambda(a,b)$, since 
we saw, in sections \ref{sec:I-zero-pt} and \ref{sec:one-point},
that the large-$n$ limit
of the zero- and one-point function in the presence of $I_\lambda(a,b)$
agrees with the corresponding twist field correlators.
The requirement that $S_\lambda(a,b)$ should have the correct monodromy
still left an $(a,b)$-dependent normalisation factor undetermined,
which is fixed in section \ref{sec:S-norm} by matching it against the leading
large-$n$ behaviour of $\langle n|I_\lambda(a,b)|0\rangle$.
Finally, in section \ref{sec:S-pole} we present the third supporting
evidence by checking, to the extend that we were able to calculate it,
that the $su(2)$-currents have the same singularities 
in the presence of $S_\lambda(a,b)$ as were seen for $I_\lambda(a,b)$
in section \ref{sec:I-mono-OPE}.

\subsection{Reproducing the monodromy of
  $I_\lambda(a,b)$}\label{sec:I-mono} 

Our first task will be to modify the product $\sigma_+(b)\sigma_-(a)$
in such a way that instead of the $\Zb_2$-monodromy, we find the
monodromy \erf{eq:I-mono} for the $su(2)$-currents. As a motivation
for the ansatz below, compare the definition \erf{eq:Idef} of
$I_\lambda(a,b)$ to the expression \erf{eq:twist-integral} for the
product $\sigma_+(b)\sigma_-(a)$. 
It seems one can go from the latter to the former by 
`subtracting' the $J^-$-contribution
to the integrals. In this spirit, define an operator
\be
  \tilde S(a,b) = C(a,b)
  \Big[ \sigma_{+\gamma}(b)
  \exp\Big( \frac{\alpha}{2\pi}
  \int_{\Cc_1} J^-(x) dx \Big) 
  \exp\Big( \frac\beta{2\pi}
  \int_{\Cc_2} J^-(x) dx \Big) 
    \sigma_{-\gamma}(a) \Big]_{\rm reg} \,.
\labl{eq:Stilde}
The contours $\Cc_{1,2}$ are as shown in \erf{eq:contour-lens},
$C(a,b)$ is a $\Cb$-valued function and $\alpha$, $\beta$, $\gamma$
are constants. Let us postpone the discussion of the regulator to section  
\ref{sec:regulate}; in any case it will be defined in such a way 
that it does not affect the monodromy.

The monodromy of the $su(2)$-currents around the point $b$ 
of $\tilde S(a,b)$ is a product of three terms, two from the
exponentiated $J^-$-integrals and one from the $\Zb_2$-monodromy
of the twist fields. In terms of the basis used in \erf{eq:I-mono}
the monodromy is
\be
  \tilde M(b) =  
  \begin{pmatrix} \! 1          & 0 & 0 \\ 
                  \! -2 i \beta \!\!\!& 1 & 0 \\ 
                  \! \beta^2    & i \beta & 1 
  \end{pmatrix} \!\!\!
  \begin{pmatrix} 0 & 0 & \gamma^{2} \\ 
                  0 & -1 \!\!\! & 0 \\ 
                  \gamma^{-2}\!\!\!\! & 0 & 0 
  \end{pmatrix} \!\!\!
  \begin{pmatrix} 1 & 0 & 0 \\ 
                  -2 i \alpha \!\!\!  & 1 & 0 \\ 
                  \alpha^2  & i \alpha & 1 
  \end{pmatrix} 
  = 
  \begin{pmatrix}
  \alpha^2 \gamma^2  &  i \alpha \gamma^2 & \gamma^2 \\
  2 i \alpha (1{-}\alpha \beta \gamma^2) &  
  2 \alpha \beta \gamma^2 -  1 
  &  -2 i \beta \gamma^2 \\ 
  \gamma^{-2} (1{-}\alpha \beta \gamma^2)^2 & 
  i \beta (\alpha \beta \gamma^2{-}1) & 
     \beta^2 \gamma^2
  \end{pmatrix} .
\labl{eq:tildeM}
We see that this matches with the monodromy $M_\lambda(b)$
of $I_\lambda(a,b)$ in (\ref{eq:I-mono}) for precisely two choices of 
the three parameters $\alpha$, $\beta$, $\gamma$, namely  
$\alpha = \beta = -\lambda^{-1}$ and $\gamma = \pm \lambda$. We will
choose $\gamma = \lambda$.

\subsection{Regulating the $J^-$-integrals}\label{sec:regulate}

To regulate the $J^-$ integrals in \erf{eq:Stilde} we will first
introduce a cutoff $\eps$ and then present a subtraction scheme which
we conjecture to give a finite $\eps\rightarrow 0$ limit.
In fact, we will regulate \erf{eq:Stilde} for the choice $\gamma=1$,
which will give the operator $S_\lambda(a,b)$ in \erf{eq:Sdef}
for $\lambda = 1$. General
values of $\lambda$ will then be obtained as in \erf{eq:Ilam-I1}.

The $\eps$-cutoff is imposed simply by changing the integration
contours $\Cc_{1,2}$ to approach the points $a$, $b$ only up
to a distance $\eps$. More precisely,
fix a small positive constant $\Lambda$ as well as a value
$\eps \ll \Lambda$. Consider the integration contours
$\tilde \Cc_{1,2}$ and $\Cc^\eps_{a,b}$ defined as follows,
\be
  \begin{picture}(200,60)(0,0)
  \put(0,0){\scalebox{0.5}{\includegraphics{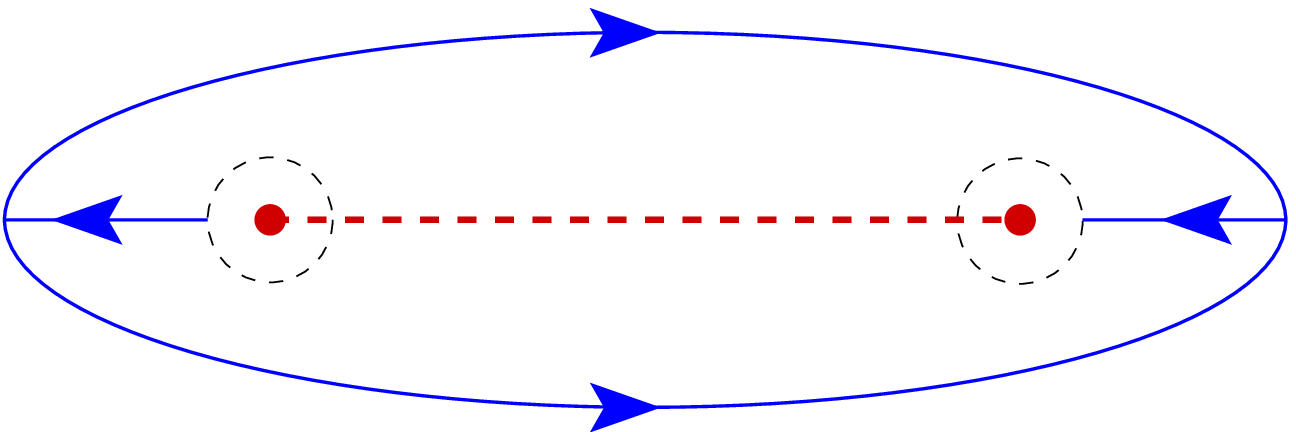}}}
  \put(45,18){$a$}
  \put(135,18){$b$}
  \put(154,55){$\tilde\Cc_1$}
  \put(154,-1){$\tilde\Cc_2$}
  \put(20,21){$\Cc_a^\eps$}
  \put(157,21){$\Cc_b^\eps$}
  \put(-22,27){$a{-}\Lambda$}
  \put(189,27){$b{+}\Lambda$}
  \end{picture}
\labl{eq:large-n-contour}
The dashed circles around the points $a$ and $b$
have radius $\eps$ so that
$\Cc^\eps_{a} = [a{-}\Lambda,a{-}\eps]$ and
$\Cc^\eps_{b} = [b{+}\eps,b{+}\Lambda]$, with orientations
as indicated. Indeed, for $\eps=0$ this is just a deformation of the
contours $\Cc_{1,2}$ defined in \erf{eq:contour-lens}.
Instead of integrating $J^-(x)$ define a field $\rho_t(x)$ as
\be
  \rho_t(x) = t J^-(x) - f(t) \langle J^-(x) \rangle \one \,,
  \qquad {\rm where} \qquad
  \langle J^-(x) \rangle \equiv 
  \frac{\langle 0| J^-(x) \sigma_+(b) \sigma_-(a) |0\rangle}{
  \langle 0| \sigma_+(b) \sigma_-(a) |0\rangle} \,.
\labl{eq:rhot-def}
Here $t$ is a complex parameter and 
$f(t) = t + f_2 t^2 + f_3 t^3 + \dots$ is a power series in $t$.
On the segments $\Cc^\eps_{a}$, $\Cc^\eps_b$ and $\tilde \Cc_{1,2}$
define the following integrated fields,
\be\bearll
  U^{\eps,\Lambda}_{+,t}(b) \!\!\! \etb
  = \exp\Big( -\frac{1}{\pi}
  \int_{\Cc_b^\eps} \rho_{t}(x) dx \Big) 
    \sigma_+(b)  \enl
  U^{\eps,\Lambda}_{-,t}(a) \!\!\! \etb
  =  
  \exp\Big( -\frac1{\pi}
  \int_{\Cc_a^\eps} \rho_{t}(x) dx \Big) 
    \sigma_-(a) 
  \enl
  V^{\Lambda}_t(a,b) & \displaystyle
  = C(a,b)
  \exp\Big( -\frac{1}{2\pi}
  \int_{\tilde\Cc_1} \rho_{t}(x) dx \Big) 
  \exp\Big( -\frac1{2\pi}
  \int_{\tilde\Cc_2} \rho_{t}(x) dx \Big) \,.
\eear\labl{eq:UV-def}
In $U^{\eps,\Lambda}_{\pm,t}$ there is a factor of $\tfrac{1}{\pi}$ instead of
$\tfrac{1}{2\pi}$ because in deforming the contour \erf{eq:contour-lens}
to \erf{eq:large-n-contour}, the segments $C^\eps_{a,b}$ are traversed twice.

The subtraction scheme now consists of expanding the operators
$U^{\eps,\Lambda}_{\pm,t}$ as a formal power series in $t$ and 
demanding that each term in the expansion has a finite limit
as $\eps \rightarrow 0$. This procedure will result in conditions 
determining the constants $f_2$, $f_3$, \dots.
It is not at all obvious that such a solution exists, 
and we have no proof that it can be done to all orders in 
$t$. In appendix \ref{sec:reg-order3} we verify that the subtraction
scheme \erf{eq:UV-def} works at least to order $t^3$, with
$f(t) = t - \tfrac{1}{\pi} t^2 + \tfrac{1}{6} t^3 + O(t^4)$.
To proceed we will assume:
\begin{itemize}
\item[(A1)] There exists a function $f(t)$ such that each order in
  the expansion of the operators $U^{\eps,\Lambda}_{+,t}(b)$
  and $U^{\eps,\Lambda}_{-,t}(a)$ in powers of $t$
  has a finite limit as $\eps\rightarrow 0$.
\end{itemize}
Using (A1), we can define
the operator $S_1(a,b)$ in terms of $U_{\pm,t}^{\eps,\Lambda}$ as
\be
  S_1(a,b) = V^\Lambda_1(a,b) ~ S^{{\rm pert}}_{+,\Lambda}(b) ~ 
  S^{{\rm pert}}_{-,\Lambda}(a) \,,
\labl{eq:Slam-S10}
where 
\be
  S^{{\rm pert}}_{+,\Lambda}(b) =
  \Big( \lim_{\eps \rightarrow 0} U^{\eps,\Lambda}_{+,t}(b)
  \Big)_{t=1} \,, \qquad
  S^{{\rm pert}}_{-,\Lambda}(a) =
  \Big( \lim_{\eps \rightarrow 0} U^{\eps,\Lambda}_{-,t}(a) \Big)_{t=1}\,.
\labl{eq:Spert-def}
Finally we then obtain
\be
S_\lambda(a,b) |0\rangle = e^{ \ln(\lambda) J^3_0 } S_1(a,b) |0\rangle
\,.
\labl{eq:Slam-S1}
A number of comments are in order. First, the abbreviation `pert' in
\erf{eq:Spert-def} stands for `perturbative', a qualifier that 
will be justified in the next section. Second, while the decomposition
\erf{eq:Slam-S10} of $S_1(a,b)$ will be useful in the following,
we can equivalently write it in a form that more closely resembles
\erf{eq:Stilde},
\be
  S_1(a,b) = C(a,b) \Bigg( \lim_{\eps\rightarrow 0}
  \sigma_+(b)
  \exp\Big( -\frac{1}{2\pi}
  \int_{\Cc_1^\eps} \rho_t(x) dx \Big) 
  \exp\Big( -\frac1{2\pi}
  \int_{\Cc_2^\eps} \rho_t(x) dx \Big) 
    \sigma_-(a) \Bigg)_{\!\!t=1} ~.
\labl{eq:S1-single-term}
Here, $\Cc_1^\eps$ is the contour obtained by joining
$\Cc_a^\eps$, $\tilde\Cc_1$ and $\Cc_b^\eps$, and 
$\Cc_2^\eps$ is obtained by joining
$\Cc_a^\eps$, $\tilde\Cc_2$ and $\Cc_b^\eps$.
In the form \erf{eq:S1-single-term} it is also apparent that
$S_1(a,b)$ has the monodromy \erf{eq:tildeM}, since
$\rho_1(x)$ differs from $J^-(x)$ only by a central term.

Third, it is clear from the definition that 
$S^{{\rm pert}}_{+,\Lambda}(b)$ and $S^{{\rm pert}}_{-,\Lambda}(a)$
are coherent states in the $\Zb_2$ twisted representations generated
by $\sigma_+$ and $\sigma_-$, respectively. 
Coherent states in twisted representations 
also appeared in relation to matrix models in 
\cite{Kostov:1999xi,Dijkgraaf:2002yn}.

\subsection{Behaviour of $S_\lambda(a,b)$ at large $n$}\label{sec:non-pert}

In section \ref{sec:one-point} we have seen that the monodromy of
$I_\lambda(a,b)$ around $a$ or $b$ is entirely due to `non-perturbative' 
effects, {\it i.e.}\ that to any finite order in $1/n$, the monodromy
is just given by the $\Zb_2$ twist \erf{eq:Z2-mono}. We now want to
show that the behaviour of $S_\lambda(a,b)$ is the same; more
explicitly, we shall show that (see \erf{eq:S-trunc}) 
\be
  \langle n | ({\rm fields}) S_\lambda(a,b) |0\rangle
  =
  \langle n | ({\rm fields}) S^{\rm trunc}_\lambda(a,b) |0\rangle
  \big( 1 + O(r^{-2n}) \big)\,,
\label{eq:S-trunc1}
\ee
for some $r>0$, where $S^{\rm trunc}_\lambda(a,b)$ has the same
monodromy as the two $\Zb_2$-twist fields. We will again only treat
the case $\lambda=1$; for the general case the reasoning is analogous
due to \erf{eq:Slam-S1}.

To define  $S^{\rm trunc}_1(a,b)$ we choose the contours 
$\tilde \Cc_{1,2}$ in \erf{eq:large-n-contour} 
to lie on the ellipse \erf{eq:modzeta-ellipse} of constant $|\zeta|$
passing through $a-\Lambda$ and $b+\Lambda$. This amounts to choosing
$r=|\zeta|$ to be
\be
  r = 1 + \tfrac{2\Lambda}{b{-}a} \Big( 1 + \sqrt{1 + 
  \tfrac{b{-}a}{\Lambda}} \, \Big) \,.
\ee
Then define the function
\be
  D^\Lambda_t(a,b) = C(a,b) \exp\!\Big( \tfrac{1}\pi f(t)
  \int_{\tilde \Cc_1} \langle J^-(x) \rangle \,dx \Big)  \,,
\ee
and set 
\be
S_1^{\rm trunc}(a,b) = D^\Lambda_1(a,b)\,
S^{{\rm pert}}_{+,\Lambda}(b) \,S^{{\rm pert}}_{-,\Lambda}(a)\,.
\ee
The claim \erf{eq:S-trunc1} is then implied by the following
statement: 
let $(F) = \prod_{i=1}^{|F|} J^{c_i}(w_i)$ abbreviate a product
of $su(2)$-currents. Then we have 
\be
  \langle n| (F) \big( V_t^\Lambda(a,b) - D^\Lambda_t(a,b) \one \big)
  \eta_+(b) \eta_-(a) |0\rangle = 
  \langle n| (F) \eta_+(b) \eta_-(a) |0\rangle \cdot
  O(r^{-2n}) \,,
\labl{eq:Vt-approx}
where $\eta_\pm(z)$ are fields corresponding to
states in the $\Zb_2$-twisted representations.
This is sufficient to establish \erf{eq:S-trunc1} since 
the fields $S^{{\rm pert}}_{\pm,\Lambda}(z)$ correspond to
(coherent) states in the $\Zb_2$-twisted representations (and thus
can play the roles of $\eta_\pm$).

To prove \erf{eq:Vt-approx}, first note that it is
enough to consider products $(F) = \prod_{i=1}^{|F|} J^{\nu_i}(w_i)$
with $\nu_i = \pm$, because $J^3$ insertions can be obtained in the
operator product expansion of $J^+$ and $J^-$. To express twist-field
descendents, define the modes $M_{r,s}$ as
\be
  M_{r,s} = \int_{\Cc_{ab}} (z{-}a)^r (z{-}b)^s J^3(z) 
\frac{dz}{2\pi i}\,,  \qquad r,s \in \Zb + \tfrac12 \,,
\ee
where $\Cc_{ab}$ is a contour encircling $a$ and $b$.
Since we are at level $k=1$, the entire $\Zb_2$-twisted representation  
is generated by acting with modes $J^3_r$, $r \in \Zb_{<0}{+}\tfrac12$ 
on $\sigma_\pm$. 
Correspondingly, the product $\eta_+(b) \eta_-(a)$ can be written
as a linear combination of appropriate products
$\prod_{i} M_{r_i,s_i} \sigma_+(a) \sigma_-(b)$, with 
$r_i{+}s_i \le 0$.
Next, note that by definition,
\be
  V_t^\Lambda(a,b) - D^\Lambda_t(a,b) \one
  = D^\Lambda_t(a,b)
  \sum_{m=1}^\infty \frac{1}{m!} \Big(\frac{-t}{2\pi}\Big)^m
   \int_{\tilde\Cc_1 + \tilde\Cc_2} \mem\mem J^-(x_1) \cdots J^-(x_m) \,
   dx_1 \cdots dx_m \,.
\labl{eq:Vt-expand}
Upon inserting \erf{eq:Vt-expand} into \erf{eq:Vt-approx} one
obtains integrands of the form
\be
  I_m = 
  \langle n| (F) J^-(x_1) \cdots J^-(x_m)
  \prod_{i=1}^M M_{r_i,s_i} |\sigma\sigma\rangle \,,
  \qquad  |\sigma\sigma\rangle = 
  \sigma_+(b) \sigma_-(a) |0\rangle \,.
\ee
Using the commutator
$[M_{r,s}, J^\pm(x)] = \pm(x{-}a)^r (x{-}b)^s J^\pm(x)$,
as well as $\langle n| M_{r,s} = \delta_{r+s,0} n  \langle n|$
(expand $M_{r,s}$ in integer modes of $J^3$ and use the condition
$r{+}s \le 0$) we
can write
\bea
  I_m = A \, \langle n| (F) J^-(x_1) \cdots J^-(x_m)
  |\sigma\sigma\rangle  \enl
  A = \prod_{i=1}^M \Big( 
  - \sum_{j=1}^{|F|} \nu_j (w_j{-}a)^{r_i} (w_j{-}b)^{s_i} 
  + \sum_{j=1}^m (x_j{-}a)^{r_i} (x_j{-}b)^{s_i} 
  + n \delta_{r_i+s_i,0} \Big) \,.
\eear\labl{eq:ImA}
To proceed we need the following two estimates on the
large $n$ behaviour of correlators,
\be
  \frac{A \, \langle n| (F) |\sigma\sigma\rangle }{
  \langle n| (F) \prod_{i=1}^M M_{r_i,s_i} |\sigma\sigma\rangle}
  = ({\rm const}) + O(n^{-1})  \,,
  \qquad
  \frac{\langle n| (F) J^-(x_1) \cdots J^-(x_m)
  |\sigma\sigma\rangle }{
  \langle n| (F) |\sigma\sigma\rangle} = O(r^{-2nm})\,.
\ee
For the first estimate note that the correlator in
the denominator produces a factor given by $A$ as in \erf{eq:ImA}
but with $m=0$. To see the second estimate, insert the explicit
solution \erf{eq:J-corr-twist} for the correlators. 
For each $J^-(x_k)$ insertion
there is a factor of $\zeta(x_k)^{-2n}$ in the numerator which,
since the $x_k$ lie on the contour $\tilde\Cc_k$, has
$|\zeta(x_k)|=r$. Then
\be\bearll
  I_m \!\!\!\etb = 
  \langle n| (F) \prod M_{r_i,s_i} |\sigma\sigma\rangle~
  \frac{A \, \langle n| (F) |\sigma\sigma\rangle }{
  \langle n| (F) \prod M_{r_i,s_i} |\sigma\sigma\rangle}~
  \frac{\langle n| (F) J^-(x_1) \cdots J^-(x_m)
  |\sigma\sigma\rangle }{
  \langle n| (F) |\sigma\sigma\rangle} 
  \enl
  \etb = \langle n| (F) \prod M_{r_i,s_i} |\sigma\sigma\rangle
    \cdot O(r^{-2nm})\,.
\eear\ee
When inserting \erf{eq:Vt-expand} into \erf{eq:Vt-approx}, the most
relevant contribution therefore comes from $m=1$ and thus is of order 
$O(r^{-2n})$. 

\subsubsection{Approximating $S_\lambda(a,b)$ in terms of twist fields}

Next we want to show that to leading order in $1/n$, $S_\lambda(a,b)$
can be replaced by a product of twist fields (see
\erf{eq:S-twist-1/n}). To obtain this relation we start
by defining a field $\rho_{t,n}(x)$ in the same way as
\erf{eq:rhot-def}, but where we subtract the one-point
function with respect to $\langle n|$ instead of $\langle 0|$,
\be
  \rho_{t,n}(x) = t J^-(x) - f(t) \langle J^-(x) \rangle_n \one \,,
  \quad {\rm where} \qquad
  \langle J^-(x) \rangle_n \equiv 
  \frac{\langle n| J^-(x) \sigma_+(b) \sigma_-(a) |0\rangle}{
  \langle n| \sigma_+(b) \sigma_-(a) |0\rangle} \,.
\ee
Let $\Cc$ be any contour from $a$ to $b$.
The difference $\rho_t(x)-\rho_{t,n}(x)$ has a well-defined integral
along $\Cc$ since the singular contribution from the poles at
$a$ and $b$ cancel.
Using \erf{eq:J-corr-twist} one finds explicitly
\bea
  \int_\Cc \big( \rho_t(x)-\rho_{t,n}(x) \big) dx
  = f(t) \int_{\Cc'} \frac{\zeta^{-2n}-1}{\zeta^2-1}
  d\zeta 
  = f(t) \sum_{l=0}^{n-1} \frac{2}{2l+1} 
  \enl \qquad
  = f(t) \big( \psi(n{+}\tfrac12) + \gamma + 2 \ln(2) \big) 
  \enl \qquad
  = f(t) \big( \ln(n) + 2 \ln(2) + \gamma + 
    \tfrac{1}{24} n^{-2} + O(n^{-4}) \big) \,,
\eear\labl{eq:rho-diff}
where $\Cc'$ is any contour from $-1$ to $1$ not passing through 
$\zeta=0$, $\gamma$ is Euler's constant and $\psi(z)$ is the digamma
function, $\psi(z) = \Gamma'(z)/\Gamma(z)$. We can then consider the
product 
\be
  \mathcal{O}_n(a,b) = 
   \exp\!\Big( \frac{1}{2\pi}
  \int_{\Cc_1+\Cc_2} \mem \big( \rho_1(x)-\rho_{1,n}(x) \big) dx \Big) \,
  S_1(a,b) \,,
\labl{eq:S1-aux1}
which amounts to replacing $\rho_1(x)$ in the 
definition \erf{eq:S1-single-term} of $S_1(a,b)$  
by $\rho_{1,n}(x)$. When using $\rho_{t,n}$
instead of $\rho_t$, both the $J^-$-integrals and the integrals of the
one-point functions $\langle J^-(x) \rangle_n$ are suppressed away
from $a$ and $b$ for large $n$. It is then plausible that in the
expansion of the exponential \erf{eq:S1-single-term},  
all terms involving $\rho_{t,n}$-integrals are of
order $O(n^{-1})$. This can be checked explicitly for the first
few terms in the expansion, but we have no general proof. 
Let us hence assume
\begin{itemize}
\item[(A2)] The operator $\mathcal{O}_n(a,b)$ in 
  \erf{eq:S1-aux1} has the large $n$ behaviour
  \be
  \langle n| ({\rm fields}) \mathcal{O}_n(a,b) |0\rangle
  =  C(a,b) \langle n| ({\rm fields}) \sigma_+(b)\sigma_-(a)|0\rangle 
  \big( 1+ O(n^{-1}) \big)\,.\labl{eq:A2}
\end{itemize}
In order to obtain a $1/n$-expansion, one should thus not consider
the correlator $\langle n|$ $({\rm fields})$ $S_1(a,b)$ $|0\rangle$  
directly, but rather the normalised version
$\langle n| ({\rm fields}) \mathcal{O}_n(a,b) |0\rangle / 
\langle n| ({\rm fields}) |\sigma\sigma\rangle$.
However, it is not true that the expansion of the exponential 
\erf{eq:S1-single-term} with $\rho_{t,n}$ instead of $\rho_t$
produces the $1/n$-expansion term by term. Instead even a 
term with $m$ $J^-$-integrations gives a contribution of
order $n^{-1}$. This is not so surprising when one considers more
carefully the regulated expression for the $m$'th term in the
expansion. In fact, the coefficient of $t^m$ will also contain a term of
the form  $f_m \int \langle J^-(x)\rangle_n dx$ from the
$t$-expansion of $\int \rho_{t,n}(x)dx$. In this
sense the subtraction scheme mixes all orders and it is
only easy to extract the large-$n$ limit, but not the subleading
terms in the $1/n$-expansion.

\subsection{Fixing the normalisation of $S_\lambda(a,b)$}\label{sec:S-norm}

In section \ref{sec:I-mono} we have partly fixed the operator 
$S_1(a,b)$ by requiring it to have the same mo\-nodromy as $I_1(a,b)$. 
Using assumption (A2) we can further fix the normalisation
$C(a,b)$ by demanding $\langle n| S_1(a,b) |0\rangle /
\langle n| I_1(a,b) |0\rangle = 1$ in the large-$n$ limit.
To this end, combine
\erf{eq:rho-diff} and \erf{eq:S1-aux1} to obtain
\be
  \mathcal{O}_n(a,b) = n^{\frac{f(1)}{\pi}} 
  e^{\frac{f(1)}{\pi}(2 \ln(2) + \gamma)} S_1(a,b) 
  \big( 1+ O(n^{-1}) \big) \,.
\ee
Next, using the above approximation together with
\erf{eq:A2} and \erf{eq:n-twist} leads to
\be
  \langle n| S_1(a,b) |0\rangle = 
  C(a,b) n^{-\frac{f(1)}{\pi}} 
  e^{-\frac{f(1)}{\pi}(2 \ln(2) + \gamma)} 4^{-n^2}
  (b-a)^{n^2-\frac18} \big( 1+ O(n^{-1}) \big) \,.
\ee
Comparing to the corresponding expression \erf{eq:n-I1} for 
$I_1(a,b)$ then leads to the requirements
\be
  f(1) = \tfrac{\pi}{4} \,,
  \qquad 
  C(a,b) = 2^{\frac{7}{12}} ~ e^{3 \zeta'(-1) + \gamma/4} ~
    (b{-}a)^{\frac18} \,.
\labl{eq:Cab}
For course, $f(1)$ is in principle determined by demanding
the $\eps$-limit in \erf{eq:S1-single-term} to exist, but one
would need the expansion to all orders to check whether its
values is indeed $\tfrac{\pi}{4}$. It may be taken as an 
encouraging sign that 
from $f(t) = t - \tfrac{1}{\pi} t^2 + \tfrac{1}{6} t^3 + O(t^4)$,
the first three approximations to
$f(1) - \tfrac{\pi}4$ are $0.21$, $-0.10$, and $0.06$.

After fixing $C(a,b)$, the operator $S_1(a,b)$ is completely
determined. The definition of the regulated expression
\erf{eq:Sdef} is that $S_1(a,b)$ is given by \erf{eq:S1-single-term} 
with $C(a,b)$ as in \erf{eq:Cab}. In \erf{eq:Sdef}, 
the constant multiplying
$(b{-}a)^{\frac18}$ in $C(a,b)$ has been absorbed into the
definition of $[\dots]_{\rm reg}$.

\subsection{Singularity structure of the $su(2)$-currents in the
presence of $S_\lambda(a,b)$}\label{sec:S-pole}

Finally we want to argue that the leading singularities as the
currents approach the endpoints of $S_1(a,b)$ are the same as for the
case of $I_1(a,b)$. For $I_1(a,b)$ these singularities were analysed
in section \ref{sec:I-mono-OPE}, and the relevant properties are
summarised in the statement (S). Here we want to present two pieces of
evidence that the conditions of the statement (S) are also met for
$S_1(a,b)$ in place of $I_1(a,b)$. In particular we want to argue that
the leading  singularity as $J^-(z)$ approaches the endpoints of 
$S_1(a,b)$ is a simple pole. We start by investigating
the behaviour of $S_1(a,b)$ under global conformal transformations.

Let $\varphi(z) = \tfrac{\alpha z + \beta}{\gamma z + \delta}$
be a M\"obius transformation and $U_\varphi$ be the operator
implementing that transformation on the space of states
(an explicit expression in terms of Virasoro generators can for
example be found in \cite[section 3.1]{Gaberdiel:1999mc}).
Denote the field $\rho_t(z)$
introduced in \erf{eq:rhot-def} by $\rho_t(z;a,b)$ to keep
track of the values for $a$ and $b$ entering its definition.
Using $U_\varphi J^-(z) U_{\varphi^{-1}} = \varphi'(z) J^-(\varphi(z))$
and $U_\varphi \sigma_\pm(z) U_{\varphi^{-1}} = 
\varphi'(z)^{\frac{1}{16}} \sigma_\pm(\varphi(z))$ it is easy to check 
that
\be
  U_\varphi \rho_t(z;a,b) U_{\varphi^{-1}} = \varphi'(z) 
  \, \rho_t\big(\varphi(z);\varphi(a),\varphi(b)\big) \,.
\ee
Next we observe that for $C(a,b)$ defined in \erf{eq:Cab}, we have 
$C(a,b) (\varphi'(a)\varphi'(b))^{\frac{1}{16}}= 
C( \varphi(a), \varphi(b))$. Thus the total effect of the M\"obius 
transformation on $S_1(a,b)$ is 
\begin{eqnarray}
U_\varphi S_1(a,b) U_{\varphi^{-1}}
& = & C(a,b) \Bigg( \lim_{\eps \rightarrow 0} 
  U_\varphi \exp\Big( -\frac{1}{2\pi} 
   \int_{\Cc_1^\eps + \Cc_2^\eps} \rho_t(x;a,b) dx \Big)
   \sigma_+(b) \sigma_-(a)
   U_{\varphi^{-1}} \Bigg)_{t=1} \nonumber \\
& = & S_1(\varphi(a),\varphi(b))\,.
\end{eqnarray}
We see that $S_1(a,b)$ transforms as a product
$\phi(a)\phi(b)$ of two Virasoro quasi-primary fields of
conformal weight zero. This observation allows us to compute
the correlator 
$\langle 0| J^c(z) S_1(a,b) |0\rangle$, for $c \in \{+,3,-\}$,
since conformal invariance
fixes a three-point function up to a constant,
\be
  \langle 0| J^c(z) S_1(a,b) |0\rangle
  = ({\rm const}) \, (z{-}a)^{-1} \, (z{-}b)^{-1} \, (a{-}b) \,.
\ee
In particular this shows that for the out-state $\langle 0|$,
all $su(2)$-currents have at most a first order pole at $a$ and $b$. 

As a second piece of evidence, we compute the correlator
$\langle n| J^\nu(z) S_1(a,b) |0\rangle$ for $\nu = \pm$
to first order in the $t$-expansion
(that is, in the definition \erf{eq:S1-single-term} of
$S_1(a,b)$ we expand the exponential to first order in $t$ 
before setting $t=1$). Instead of using
$S_1(a,b)$ it is convenient to use $\mathcal{O}_n(a,b)$
as defined in \erf{eq:S1-aux1} which differs from $S_1(a,b)$ by
a constant. One finds, with $\zeta = \zeta(z)$ and $\xi = \zeta(x)$,
\bea
  \langle n| J^\nu(z) \mathcal{O}_n(a,b) |0\rangle
 \enl
  = C(a,b) \Bigg(
  \langle n| J^\nu(z) \Big( 1 - \frac{t}{2\pi}
  \int_{\Cc_1+\Cc_2} \big( J^-(x) - \langle J^-(x) \rangle_n \big) dx
  + O(t^2) \Big) |\sigma\sigma\rangle \Bigg)_{t=1} 
\enl
  = C(a,b)  \langle n| J^\nu(z)  |\sigma\sigma\rangle \Bigg(
   1 + \frac{t}{2\pi} (\zeta^{2\nu}-1)
  \int_{\Cc_1'+\Cc_2'} \frac{\xi^{-2 n}}{(\xi \zeta^\nu -1)^2} d\xi
  + O(t^2)  \Bigg)_{t=1}  \,.
\eear\labl{eq:pole-aux1}
Here $\Cc_1'$ is a contour from $-1$ to $1$ passing along the
upper half of the unit circle, while $\Cc_2'$ passes along the lower
half of the unit circle. The integral in \erf{eq:pole-aux1} 
is given by
\be
  \int_{\Cc'} \frac{x^{-2n}}{(xy-1)^2} \, dx = \Big[- \frac{1}{2n+1}\,
  \frac{x^{-2n+1}}{(xy-1)^2} \, {}_2F_1\big(1,2;2n+2;(1-xy)^{-1}\big) 
  \Big]_{x=-1}^{x=1} \,,
\labl{eq:pole-aux2}
where $\Cc'$ is a contour from $-1$ to $1$, which also determines
the relevant branch of the hypergeometric function.
We are interested in the asymptotics of \erf{eq:pole-aux1} for
$\zeta \rightarrow \pm 1$. This amounts to taking the argument
of the hypergeometric function in \erf{eq:pole-aux2} to infinity.
The hypergeometric function has the asymptotics, for
$u\rightarrow\infty$, 
\be
  {}_2F_1(1,2;2n{+}2;u) = - (2n{+}1)\,u^{-1} + 
  O\big( u^{-2} \ln(u) \big) \,.
\ee 
Altogether we find that the leading singularities of the integral
\erf{eq:pole-aux2} are first order poles at $y=\pm 1$, which in
\erf{eq:pole-aux1} get cancelled by the prefactor $\zeta^{2\nu}{-}1$.
Hence to first order in $t$, the leading singularity of 
\erf{eq:pole-aux1} for $z \rightarrow a,b$ is that of 
$\langle n| J^\nu(z)  |\sigma\sigma\rangle$, which clearly satisfies
the conditions of (S).
\medskip

After presenting these two calculations regarding the poles
of $J^c(z)S_1(a,b)$ for $z \rightarrow a,b$, we will assume that
in general
\begin{itemize}
\item[(A3)] for $x=a,b$, inside any correlator,
$\lim_{z\rightarrow x} (z-x) J^-(z) S_1(a,b)$  
is finite, and 
there exists some $N > 0$ such that
$\lim_{z\rightarrow x} (z-x)^N J^{+,3}(z) S_1(a,b)$ is zero.
\end{itemize}
It then follows that the fields at the endpoints of $S_1(a,b)$, just
as for $I_1(a,b)$, are highest weight with respect to the single
valued combinations \erf{eq:hat-J}, and that in particular they are
Virasoro primary of weight zero, {\it i.e.} 
$L_{m;a} \, S_1(a,b) = 0 = L_{m;b} \, S_1(a,b)$ for $m \ge 0$.
Together with the fact that $I_\lambda(a,b)$ and $S_\lambda(a,b)$ have
the same monodromy properties, this is 
very good evidence for the
equality of $I_\lambda(a,b)$ and $S_\lambda(a,b)$.

\sect{Outlook}\label{sec:Out}

In this paper we have proposed and provided evidence for
the operator identity $I_\lambda(a,b) = S_\lambda(a,b)$.
Here $I_\lambda(a,b)$ has a simple formulation as an exponentiated
integral of $J^+$-currents and is directly related to eigenvalue
integrals in matrix models. The expression for
$S_\lambda(a,b)$ is more complicated, involving twist fields and a
regulator. However, properties of the large-$n$ limit of correlators
are easily understood in terms of $S_\lambda(a,b)$ while they are
harder to see when using $I_\lambda(a,b)$.

\medskip\noindent
There are several directions in which one can attempt to generalise
the analysis of this paper.
\\[.3em]
(i) From the conformal field theory point of view, the most obvious
question is whether there is a generalisation to $su(2)$ at level
$k>1$. In this case the operator $I_\lambda(a,b)$ is still given by
\erf{eq:Idef}. However, there are now $k{+}1$ $\Zb_2$-twisted highest
weight representations, and there is no longer a unique (up to scalar 
multiples) out-state 
$\langle n|$ which is highest weight with respect to 
$J^3_m$ and has $J^3$-charge $n$, but a finite-dimensional subspace.
Both effects make it more difficult to identify the analogue of
$S_\lambda(a,b)$. Independently, it is also of interest to see if 
correlators of $I_\lambda(a,b)$ in $su(2)_k$ do have a matrix model
interpretation.
\\[.3em]
(ii) Recently a conformal field theory approach to the non-linear
$\sigma$ model with the complex torus $(C^*)^d$ as target space has been  
investigated in \cite{Frenkel:2005ku}. It uses a non-unitary CFT 
which contains a $su(2)$ algebra at level $k=0$. Similar to
(\ref{eq:Idef})  
it has an integrated exponential operator, which does not 
require normal ordering and generates a logarithmic branch cut. 
The above model is related to the A-model by a deformation 
and to the B-model by an additional T-duality.
It would be very interesting to understand whether this CFT approach
to the non-linear $\sigma$-model is linked to a matrix model
description.  On non-compact Calabi-Yau 3-folds associated to ADE 
singularities~\cite{Dijkgraaf:2002vw} the topological B-model, a subsector 
of the non-linear $\sigma$-model, has already been related to ADE-quiver 
matrix models \cite{Dijkgraaf:2002fc,Dijkgraaf:2002vw,Chiantese:2003qb}, 
see also (iv) below.       
\\[.3em]
(iii) From the matrix model point of view, one should use more general
potentials than the infinite well potential that we considered here.
This amounts to taking the out-state $\langle n| e^{-H}$ instead of
just $\langle n|$. Also, it would be good 
to remove the effect of the hard edges, possibly by considering
$S_\lambda(a-\eps_n,b+\eps_n)$ to shift the endpoints
by an $n$-dependent amount away from the location of a cut.
\\[.3em]
(iv) There is a whole class of multi-matrix models, called 
ADE-quiver matrix models, which can be rewritten in terms of free
bosons, or, more precisely, in terms of an ADE-WZW model at level one
\cite{Kostov:1992ie,Kharchev:1992iv,Morozov:hh,Kostov:1995xw,
Dijkgraaf:2002vw,Chiantese:2003qb}. For these one would define
a number of operators $I_\lambda^i(a,b)$,  
indexed by simple roots $\alpha_i$, $i=1,\dots,r$. Since each of the
corresponding raising operators $E^{\alpha_i}$ lies in an
$su(2)$-subalgebra, one would expect that the analysis of this paper
can be repeated without much modification. The comparison of
perturbative versus non-perturbative moduli spaces of FZZT-branes (as
mentioned in the introduction) has also been carried out for
$(p,1)$-minimal string theory using a two-matrix model
\cite{Hashimoto:2005bf}  (however, not an ADE-quiver 
model) and it would be interesting to compare results.
\\[.3em]
(v) Insertions of several operators $I_{\lambda_1}(a_1,b_1)$,
$I_{\lambda_2}(a_2,b_2)$, \dots correspond to several cuts in 
the complex plane. On the matrix model side one obtains in this
way not a partition function with fixed filling fractions, but
rather a generating function in the parameters $\lambda_k$ for
the various filling fractions. It would be interesting to see
if the methods of this paper can be extended to be a useful tool in
investigating the $1/n$ expansion of such multi-cut solutions. 
\\[.3em]
We hope to address some of these points in the future.

\bigskip
\section*{Acknowledgements}
AK is supported in parts by DOE grant DE-FG02-95ER40896.
The research of MRG has been partially supported by the Swiss
National Science Foundation and the Marie Curie network `Constituents,
Fundamental Forces and Symmetries of the Universe'
(MRTN-CT-2004-005104). IR thanks the organisers of the workshop
`Random Matrices and Other Random Objects' (Z\"urich, May 2005) for an
interesting and stimulating meeting. We acknowledge useful
conversations with Stefano Chiantese, Bertrand Eynard, Kevin Graham,
J\"urg Fr\"ohlich, Sebastian de Haro, Ivan Kostov, Matthias
Staudacher, Miguel Tierz and Paul Wiegmann.

\appendix

\sect{Appendix}

\subsection{Calculation of $J^3$ and $J^-$ pole with $I_\lambda(a,b)$}
\label{sec:I-pole-cond}

In this appendix we show that the $J^3$- and $J^-$-conditions of 
statement (S) in section \ref{sec:I-mono-OPE} are met. For the case of 
$J^-$ we need to prove that 
\be\label{A.1}
\lim_{z\rightarrow b} \,(z-b) \, 
\big\langle\prod_j J^{a_j}(w_j) \, J^-(z) I_\lambda(a,b) \,\big\rangle = 
{\rm finite}  \,, 
\ee
where the $w_j\ne a,b$ are pairwise disjoint. Then 
\be\label{A.2}
\big\langle \prod_j J^{a_j}(w_j)\,
J^-(z) I_\lambda(a,b) \,\big\rangle = 
\frac{1}{n!}\, \left(\frac{\lambda}{2\pi}\right)^{n} \,
\int_a^b \!\! dz_1 \, \cdots \int_a^b \!\! dz_{n} \,
\big\langle \prod_j J^{a_j}(w_j)\,
J^-(z) J^+(z_1) \cdots J^+(z_{n}) \,\big\rangle  
\ee
for some suitable $n$. The amplitude in the integrand can be
calculated recursively, using the singular part of the operator
product expansion (see for example \cite{Gaberdiel:1999mc}). Starting
with the $J^+$ fields, it is then obvious that the terms that could 
be singular in the limit $z\rightarrow b$ arise in two ways: either we
have the double pole  
\be
J^+(z_i) J^-(z) \sim \frac{1}{(z_i-z)^2} 
\ee
that gives rise, after integration of $z_i$, to a simple pole in 
$(z-b)$; this contribution is therefore finite in the limit of
(\ref{A.1}). The second contribution comes from the simple pole 
\be
J^+(z_i) J^-(z) \sim \frac{2 J^3(z)}{z_i-z} \,.
\ee
The other $J^+(z_j)$ fields can then either contract with the 
$J^{a_j}(w_j)$ fields, or we can get a further contraction of
$J^+(z_j)$ with $J^3(z)$, which then leads to 
\be
J^+(z_j) J^+(z_i) J^-(z) \sim - \frac{2 J^+(z)}{(z-z_i)(z-z_j)} \,.
\ee
In either case, it is straightforward to determine the $z_i$
integrals, and we obtain either a $\log(z-b)$ or a $\log^2(z-b)$
term. In the limit of (\ref{A.1}) these contributions therefore
vanish. 

The analysis for the case of $J^3$ is essentially identical; in this
case, only the second type of terms appear, and we find with the same
arguments as above that 
\be
\lim_{z\rightarrow b} \,(z-b) \, 
\big\langle\prod_j J^{a_j}(w_j) \, J^3(z) I_\lambda(a,b) 
\,\big\rangle = 0 \,.
\ee

\subsection{Existence of the regulator to order $t^3$}\label{sec:reg-order3}

In this section we compute the first few orders in $t$ of the 
function $f(t) = f_1 t + f_2 t^2 + f_3 t^3 + O(t^4)$
which enters the definition \erf{eq:Slam-S10} of $S_1(a,b)$
via \erf{eq:rhot-def}. To this end we make use of the decomposition
\erf{eq:Slam-S10} of $S_1(a,b)$. 
The operators $S^{{\rm pert}}_{\pm,\Lambda}$ are well defined if
$U^{\eps,\Lambda}_{\pm,t}$ (as given in \erf{eq:UV-def}) has a 
finite $\eps\rightarrow 0$ limit. This requirement
fixes the constants $f_1$, $f_2$, $f_3$ uniquely. Here we will only
treat $S^{{\rm pert}}_{+,\Lambda}(b)$ explicitly. The calculation 
for $S^{{\rm pert}}_{-,\Lambda}(a)$ is analogous and leads to the
same answer.

In section \ref{sec:S-pole} it was shown that under global conformal
transformations, $S_1(a,b)$ behaves as a product $\phi(a)\phi(b)$ of
Virasoro-primary fields of weight zero. We will use this freedom to
assume that $b=0$ and $a=-\infty$. The question whether 
$S^{{\rm pert}}_{+,\Lambda}(b)$ is well-defined now amounts  
to checking that
\be
   U^{\eps,\Lambda}_{+,t}(0)|0 \rangle = 
   \exp\Big( \frac1{\pi}
  \int_{\eps}^\Lambda \mem\rho_{t}(x) \, dx \Big) |\sigma_+\rangle
\labl{eq:ueps-def}
has a finite $\eps\rightarrow 0$ limit, order by order in $t$
(the sign difference with respect to \erf{eq:UV-def} is due to the
change of direction in the integral). Here, $\rho_t(x)$ takes the form
\be
  \rho_t(x) = t J^-(x) - f(t) \lim_{a\rightarrow -\infty}
  \frac{\langle 0| J^-(x) \sigma_+(0) \sigma_-(a) \rangle}{
  \langle 0| \sigma_+(0) \sigma_-(a) \rangle} 
  = t J^-(x) - \frac{f(t)}{4 x} \,.
\ee
Set further $R(x) = J^-(x) - f_1/(4x)$. Then the first few orders in
the $t$-expansion of \erf{eq:ueps-def} read
\be\bearll
  U^{\eps,\Lambda}_{+,t}(0)|0 \rangle = \etb \!\!\! |\sigma_+\rangle
  + t  \frac{1}{\pi} \int \! R(x) \,dx\, |\sigma_+\rangle
  + t^2 \Big( \frac{1}{\pi^2}  \int_{x>y} \mem  R(x) R(y) \, dx \, dy 
    - \frac{1}{\pi} \int \! \frac{f_2}{4x} \, dx 
    \Big)|\sigma_+\rangle
  \enl    
  \etb +~ t^3 \Big( \frac{1}{\pi^3}  \int_{x>y>z} \mem\mem 
    R(x) R(y) R(z) \,dx\,dy\,dz 
    - \frac{1}{\pi^2} \int_{x>y} \mem R(x) \frac{f_2}{4y} \,dx \,dy 
\enl
   \etb \qquad\qquad -~ \frac{1}{\pi^2} \int_{x>y} \frac{f_2}{4x} R(y) \,dx\, dy 
    - \frac{1}{\pi} \int \!\frac{f_3}{4x} \,dx 
    \Big)|\sigma_+\rangle
  ~+~ O(t^4) \,,
\eear\labl{eq:ueps-expand}
where the integrations are from $\eps$ to $\Lambda$, subject to the
path ordering constraints as indicated. 

\subsection*{Order $t$}

Our strategy will be to expand $R(x)$ in modes around zero and analyse
which modes give divergent contributions to the integral. To do so,
first express $R(x)$ in the $K$-basis \erf{eq:K1-basis},
\be
  R(x) = K^3(x) - \frac{1}{2}\Big( K^+(x) - K^-(x) \Big) 
- \frac{f_1}{4x} \,.
\labl{eq:R-via-K}
The field $K^3(x)$ has integer modes and $K^\pm(x)$ half-integer
modes. We decompose
\be\begin{array}{lll}
  K^3(x) = K^3_{\ge}(x) + K^3_<(x)\,, 
  \etb K^3_{\ge}(x) = \sum_{m \in\Zb_{\ge 0}} x^{-m-1} K^3_m \,,
  \etb K^3_{<}(x) = \sum_{m \in\Zb_{<0}} x^{-m-1} K^3_m \,,
  \enl
  K^\pm(x) = K^\pm_{>}(x) + K^\pm_<(x) \,,
  \etb K^\pm_{>}(x) = \sum_{r \in\Zb_{\ge 0} 
           + \frac12} x^{-r-1} K^\pm_r ,
  \etb K^\pm_{<}(x) = \sum_{r \in\Zb_{<0}+ \frac12} x^{-r-1} K^\pm_r \,.
\eear\ee
We also set $R_<(x)$ and $R_{\ge}(x)$ 
to be be given by \erf{eq:R-via-K} with all fields
replaced by their $<$-part, respectively their $\ge$- or $>$-part;
the $\tfrac{f_1}{4x}$ term is part of $R_{\ge}(x)$.
The coefficient of $t$ in \erf{eq:ueps-expand} becomes
\be
\frac{1}{\pi} \int \! R(x) \, dx \, |\sigma_+\rangle = 
\frac{1}{\pi} \int \! R_<(x) \,dx \,|\sigma_+\rangle
+ \frac{1}{\pi}\int \!\big( K^3_0 - \tfrac14 f_1 \big) 
x^{-1} dx \,|\sigma_+\rangle \,.
\ee
The integral over $R_<(x)$ involves only powers $x^r$ with
$r \ge -\tfrac12$ and has a finite $\eps\rightarrow 0$ limit. The
second integral has a log-divergence for $\eps\rightarrow 0$ which
has to be cancelled. This is achieved by setting $f_1=1$. With this
choice for $f_1$ we have 
\be
  R_{\ge}(x) \, |\sigma_+\rangle = 0 \,.
\labl{eq:Rge=0}

\subsection*{Order $t^2$}

For the second order computation, we need to know the commutators
$[K^a_{\ge}(x), K^b(y)]$ for $x>y$. These can be computed from the
commutation relations of the $K^a_m$-modes, which are just the same
as those of the $J^a_m$-modes given in \erf{su2comm}. One finds,
for $\nu = \pm$,
\be\begin{array}{ll}\displaystyle
\big[ \, K^3_{\ge}(x) \,,\, K^3(y) \,\big] 
  \etb = ~ \frac{1}{2(x-y)^2} \enl
\big[ \, K^3_{\ge}(x) \,,\, K^\nu(y) \,\big] 
  \etb = ~ \frac{\nu}{x-y} ~ K^\nu(y) \enl
\big[ \, K^\nu_{>}(x) \,,\, K^3(y) \,\big] 
  \etb = ~ - \sqrt{\frac{y}{x}} ~ \frac{\nu}{x-y} ~ K^\nu(y) \enl
\big[ \, K^\nu_{>}(x) \,,\, K^{-\nu}(y) \,\big] 
  \etb = ~ \frac{1}{2(x-y)^2}
  \Big( \sqrt{\frac{y}{x}} + \sqrt{\frac{x}{y}} \Big)
  + \sqrt{\frac{y}{x}}~\frac{2\nu}{x-y} ~ K^3(y)  \,.
\eear\labl{eq:Kpos-K-com}
For the coefficient of $t^2$ in \erf{eq:ueps-expand} we then obtain
\bea
\Big( \frac{1}{\pi^2}  \int_{x>y} \mem R(x) R(y) \, dx \, dy 
    - \frac{1}{\pi} \int \frac{f_2}{4x} \, dx 
    \Big)|\sigma_+\rangle
\enl    
\overset{1)}{=} 
\Big( \frac{1}{\pi^2}  \int_{x>y} \mem R_<(x) R_<(y) \, dx \, dy 
    + \frac{1}{\pi^2}  \int_{x>y} \mem \big[R_{\ge}(x) , R(y)\big] \, dx \, dy 
    - \frac{f_2}{4\pi} (\ln\Lambda - \ln\eps)  
    \Big)|\sigma_+\rangle
\enl
\overset{2)}{=} 
\Big( -\frac{1}{2 \pi^2}  \int_{x>y}  
  \frac{K^+(y) + K^-(y) }{\sqrt{x}(\sqrt{x}+\sqrt{y})}
         \, dx \, dy 
  - \frac{1}{4\pi^2}  \int_{x>y}  \frac{1}{\sqrt{xy} 
         (\sqrt{x} + \sqrt{y})^2} \, dx \, dy 
\enl\hspace*{5em}         
    + \frac{f_2}{4\pi} \ln\eps + ({\rm finite~}\eps{\rightarrow}0)  
    \Big)|\sigma_+\rangle
\enl    
\overset{3)}{=} 
\Big(\frac{1}{4\pi^2} \ln\eps 
    + \frac{f_2}{4\pi} \ln\eps + ({\rm finite~}\eps{\rightarrow}0)  
    \Big)|\sigma_+\rangle \,,
\eear\ee
where in step 1) we replaced $R(x)$ by $R_<(x) + R_{\ge}(x)$ and
used \erf{eq:Rge=0}.
For step 2) note that the integral over $R_<(x) R_<(y)$ has a
finite $\eps\rightarrow 0$ limit. The commutator
$[R_{\ge}(x) , R(y)]$ has been evaluated with the help of
\erf{eq:Kpos-K-com}. 
For step 3), the singular contributions to the resulting integrals
have to be extracted. 
The most singular term in the first integrand
(expand $K^\nu(y)$ in modes) is 
$( \sqrt{xy}(\sqrt{x} + \sqrt{y}) )^{-1}
= 4 \partial_x \partial_y( (\sqrt{x}+\sqrt{y})\ln(\sqrt{x}+\sqrt{y}))$, 
so that the integral is regular for $\eps{\rightarrow}0$. 
The integrand of the second integral is
$-4 \partial_x \partial_y \ln(\sqrt{x}+\sqrt{y})$ so that there
is a $\ln\eps$ singularity.
In order to have a finite $\eps{\rightarrow}0$ limit at order
$t^2$ we need to choose $f_2=-\pi^{-1}$.

\subsection*{Order $t^3$}

The coefficient of $t^3$ in \erf{eq:ueps-expand} can be
written in the form $A_1 + A_2 + A_3$ with
\bea
  A_1 = \frac{1}{\pi^3}  \int_{x>y>z} \mem\mem  R(x) R(y) R(z) 
  \, dx \, dy \, dz  \, |\sigma_+\rangle
\enl
  A_2 = - \frac{f_2}{4\pi^2} \int_{x>y} \! \Big( 
     \frac{R(x)}{y} + \frac{R(y)}{x} \Big) \, dx \, dy \, |\sigma_+\rangle
\enl
  A_3 =  - \frac{f_3}{4\pi} \int \frac{1}{x} \, dx \, |\sigma_+\rangle \,.
\eear\ee
The singular contributions to $A_2$ and $A_3$ are easy to evaluate,
\be
  A_2 = \ln(\eps) \frac{f_2}{4 \pi^2}  \int \! R_<(x) \,dx\, |\sigma_+\rangle
  + ({\rm finite}~\eps{\rightarrow}0) \,,
  \qquad
  A_3 = \ln(\eps) \frac{f_3}{4\pi} |\sigma_+\rangle
  + ({\rm finite}~\eps{\rightarrow}0) \,.
\ee
Extracting the singular part of $A_1$ is some work. 
One first uses the commutators
\erf{eq:Kpos-K-com} to remove all positive mode parts of the
fields $K^a$. One obtains a sum of terms, with each term
being a function in $x,y,z$ multiplying a state of the form,
for $u,v \in {x,y,z}$,
\be
 |\sigma_+\rangle \,, \qquad
 K^a_{<}(u)|\sigma_+\rangle \,, \qquad
 K^a_{<}(u)K^b_{<}(v)|\sigma_+\rangle \,, \qquad
 K^a_{<}(u)K^b_{<}(v)K^c_{<}(u)|\sigma_+\rangle \,. 
\ee
It turns out that only the coefficients of the first two
terms lead to singular behaviour as $\eps \rightarrow 0$.
In fact on finds
\be
  A_1
  = \frac{1}{4\pi^3} I |\sigma_+\rangle
    - \frac{1}{8 \pi^3} \int_\eps^\Lambda \int_\eps^\Lambda 
         \int_\eps^\Lambda \frac{R_<(x)|\sigma_+\rangle}{
        \sqrt{y z}(\sqrt{y}+\sqrt{z})^2} dx\, dy\, dz
        + ({\rm finite}~\eps{\rightarrow}0) \,,
\ee
where
\be
  I = \int_{x>y>z} \frac{dx\, dy\, dz}{\sqrt{xyz}(\sqrt{x}+\sqrt{y})
     (\sqrt{x}+\sqrt{z})(\sqrt{y}+\sqrt{z})} 
     = -\frac{\pi^2}6 \ln(\eps) + ({\rm finite}~\eps{\rightarrow}0) \,.
\ee
Altogether
\bea
  A_1 + A_2 + A_3 
  \\[.1em]\displaystyle
  = \ln(\eps) \Big( -\frac{1}{24\pi} 
  + \frac{1}{4 \pi^3} \int_\eps^\Lambda \mem R_<(x) \, dx  
  + \frac{f_2}{4 \pi^2} \int_\eps^\Lambda \mem R_<(x) \, dx  
  + \frac{f_3}{4\pi} \Big) |\sigma_+\rangle
  + ({\rm finite}~\eps{\rightarrow}0) \,,
\eear\ee
which has a finite limit iff $f_2=-\pi^{-1}$ and $f_3=\tfrac16$. Note that
the required value for $f_2$ agrees with the one obtained in the order
$t^2$ computation.

\raggedright

\end{document}